\documentclass[final,5p,11pt,times,twocolumn]{elsarticle}
 
\usepackage{amsmath}

\usepackage{amssymb}
\usepackage{graphicx}
\usepackage{hyperref,cleveref}
\hypersetup{colorlinks=true,linkcolor=blue,urlcolor=blue,citecolor=blue}

\usepackage{wrapfig}
\usepackage{comment}
\usepackage{multirow}
\usepackage{booktabs}
\usepackage[table,xcdraw]{xcolor}
\usepackage[para,online,flushleft]{threeparttable}
\usepackage{siunitx}
\usepackage[table,xcdraw]{xcolor}
\usepackage{arydshln}
\usepackage{color, colortbl}
\usepackage[normalem]{ulem} 
\definecolor{Gray}{gray}{0.93} 
\definecolor{turquoise}{RGB}{225,255,254}
\definecolor{magenta}{RGB}{255, 0, 151}
\definecolor{lightGreen}{RGB}{144, 200, 134}
\definecolor{myblue}{RGB}{0, 92, 202}

\usepackage{xcolor}
\usepackage{pdflscape,multicol,blindtext}
\usepackage{rotating}
\usepackage{afterpage}
\usepackage{lscape}

\makeatletter\newcommand*\mysizea{\@setfontsize\mysizea{9.5}{11}}\makeatother
\makeatletter\newcommand*\mysizeb{\@setfontsize\mysizeb{8.25}{11}}\makeatother
\makeatletter\newcommand*\mysizec{\@setfontsize\mysizeb{7.15}{11}}\makeatother

\journal{Computer Methods and Programs in Biomedicine}

\begin{document}
\begin{frontmatter}

\title{Open-Full-Jaw: An open-access dataset and pipeline\\ for finite element models of human jaw}

\author[inst1,inst2]{Torkan Gholamalizadeh \corref{cor1}} 
\author[inst1]{Faezeh Moshfeghifar}
\author[inst3]{Zachary Ferguson}
\author[inst4]{Teseo Schneider}
\author[inst3]{Daniele Panozzo}
\author[inst1]{Sune Darkner}
\author[inst5,inst6]{Masrour Makaremi}
\author[inst6]{François Chan}
\author[inst2]{Peter Lempel Søndergaard}
\author[inst1]{Kenny Erleben}
\cortext[cor1]{Corresponding author: torkan@di.ku.dk}

\affiliation[inst1]{organization={Department of Computer Science, University of Copenhagen},
            city={Copenhagen},
            postcode={2100},
            country={Denmark}
            }

\affiliation[inst2]{organization={3Shape A/S},
            city={Copenhagen},
            postcode={1060},
            country={Denmark}}
            
\affiliation[inst3]{organization={Courant Institute of Mathematical Sciences, New York University},
            city={60 5th Ave, New York, NY 10011},
            country={United States}}

\affiliation[inst4]{organization={Department of Computer Science, University of Victoria},
            city={Victoria, BC V8P 5C2},
            country={Canada}}  
            
\affiliation[inst5]{organization={Dentofacial Orthopedics Department, University of Bordeaux},
            city={Bordeaux},
            country={France}}
            
\affiliation[inst6]{organization={Orthodontie clinic},
            city={2 Rue des 2 Conils, 24100 Bergerac},
            country={France}}

\begin{abstract} 
\paragraph{Background} State-of-the-art finite element studies on human jaws are mostly limited to the geometry of a single patient.
In general, developing accurate patient-specific computational models of the human jaw acquired from cone-beam computed tomography (CBCT) scans is labor-intensive and non-trivial, which involves time-consuming human-in-the-loop procedures, such as segmentation, geometry reconstruction, and re-meshing tasks. 
Therefore, with the current practice, researchers need to spend considerable time and effort to produce finite element models (FEMs) to get to the point where they can use the models to answer clinically-interesting questions. Besides, any manual task involved in the process makes it difficult for the researchers to reproduce identical models generated in the literature. Hence, a quantitative comparison is not attainable due to the lack of surface/volumetric meshes and FEMs.
\paragraph{Methods}
We share an open-access repository composed of 17 patient-specific computational models of human jaws and the utilized pipeline for generating them for reproducibility of our work.
The used pipeline minimizes the required time for processing and any potential biases in the model generation process caused by human intervention. 
It gets the segmented geometries with irregular and dense surface meshes and provides reduced, adaptive, watertight, and conformal surface/volumetric meshes, which can directly be used in finite element (FE) analysis. 

\paragraph{Results}
We have quantified the variability of our 17 models and assessed the accuracy of the developed models from three different aspects; (1) the maximum deviations from the input meshes using the Hausdorff distance as an error measurement, (2) the quality of the developed volumetric meshes, and (3) the stability of the FE models under two different scenarios of tipping and biting.

\paragraph{Conclusions}
The obtained results indicate that the developed computational models are precise, and they consist of quality meshes suitable for various FE scenarios. We believe the provided dataset of models including a high geometrical variation obtained from 17 different models will pave the way for population studies focusing on the biomechanical behavior of human jaws.
\end{abstract}




\end{frontmatter}


\section{Introduction}

Finite element modeling (FEM) is a numerical approach for predicting responses of different tissues under physical loads, which can be difficult or impossible to measure directly in vivo \cite{ausiello2020stress}. It is a widely used tool as a pre-operative protocol in different medical applications such as orthopedic surgery, orthodontic treatments, and cardiovascular surgeries \cite{benaissa2020stress}. More specifically, in the orthodontic and dental fields of studies, FEM is utilized to predict teeth movements, stress/strain distribution in different tissues (e.g., periodontal ligament, gingiva, and alveolar bone) or orthodontic appliances \cite{cattaneo2021orthodontic}. Except for a few recent studies \cite{gholamalizadeh2020mandibular,gholamalizadeh2021multi,savignano2020three}, almost all of the previous studies in the field are limited to single-patient analysis \cite{seo2021comparative,ding2014influence,ortun2020silico,benaissa2020stress,vukicevic2021openmandible,boryor2009downloadable}, in which the results might not be generalized to a larger population with high geometrical variations in the teeth, periodontal ligament (PDL), and bone anatomies \cite{gholamalizadeh2020mandibular,savignano2020three}.

\begin{table*}[t]
\caption{Related studies in the literature and their details on the data availability and mesh conformity. The few studies with public models are listed below the dashed line. Note that there are empty cells in some studies, as it was difficult to evaluate the meshes.}
\vspace{1.5mm}
\resizebox{\textwidth}{!}{%
\begin{threeparttable}
\begin{tabular}{lccllllccccc}
\hline
\multicolumn{1}{c}{\multirow{2}{*}{Study/Dataset}} & \multicolumn{2}{c}{Cohort Info} & \multicolumn{1}{c}{Scan Info} & \multicolumn{3}{c}{Geometry and Discretization Info} & \multicolumn{2}{c}{Interface info} & \multicolumn{3}{c}{Availability} \\ \cline{2-12} 
\multicolumn{1}{c}{} & \#Patient & \#Jaws & \multicolumn{1}{c}{Scan Modality} & \multicolumn{1}{c}{\begin{tabular}[c]{@{}c@{}}Tooth\\ Type\end{tabular}} & \multicolumn{1}{c}{\begin{tabular}[c]{@{}c@{}}PDL\\ Type\end{tabular}} & \multicolumn{1}{c}{\begin{tabular}[c]{@{}c@{}}Bone\\ Type\end{tabular}} & \begin{tabular}[c]{@{}c@{}}Interface\\  Congruency\end{tabular} & \begin{tabular}[c]{@{}c@{}}Mesh\\ Conformity\end{tabular} & \begin{tabular}[c]{@{}c@{}}Surface\\ Meshes\end{tabular} & \begin{tabular}[c]{@{}c@{}}Volumetric \\ Meshes\end{tabular} & \begin{tabular}[c]{@{}c@{}}FE model \\ file\end{tabular} \\ \hline
Seo et al. (2021) \cite{seo2021comparative} & 1 & Half-arch & CBCT & H, Uniform & H, Uniform & CC, Uniform & CG & CM & No & No & No \\ 
Ding et al. (2014) \cite{ding2014influence} & 1  & 1 Mandible & CBCT + MRI & EDP, Adaptive & H, Uniform & CC, Semi-Adaptive &&& No & No & No \\
Savignano et al. (2020) \cite{savignano2020three} & 2 & 4 Jaws & CBCT & H, Uniform & H, Uniform & H, Uniform & CG & Non-CM & No & No & No\\
Ortun et al. (2020) \cite{ortun2020silico}& 1 & 2 Jaws & CBCT + $\mu$CT & EDP, Uniform & H, Uniform & CC, Uniform & CG & Non-CM & No & No & No\\
Benaissa et al. (2020) \cite{benaissa2020stress} & 1 & Single Tooth & CT & H, Adaptive & H, Uniform & CC, Adaptive & & Non-CM & No & No & No \\
Kawamura et al. (2019) \cite{kawamura2019biomechanical} & DM & 1 Mandible & N/A & H, Uniform & H, Uniform & N/A & ~ CG $^{\dagger}$ & CM  & No & No & No \\
Kawamura et al. (2022) \cite{kawamura2022biomechanical} & DM & 1 Maxilla & N/A & H, Uniform & H, Uniform & N/A & ~ CG $^{\dagger}$ & CM & No & No & No \\
Oenning et al. (2018) \cite{oenning2018resorptive} & 1 & Half-arch & CBCT & EDP, Uniform & H, Uniform & CC & Non-CG &  & No & No & No\\
Sarrafpour et al. (2013) \cite{sarrafpour2013tooth} & ~~1$^{\star}$ & 1 Mandible & CBCT & EDP, Uniform & H, Uniform & CC, Adaptive & ~~Non-CG $^{\dagger}$ & Non-CM & No & No & No\\
Lee et al. (2018) \cite{lee2018biomechanical} & 1 & 2 jaws & CBCT & EDP & H & CC, Uniform & CG & Non-CM & No & No & No\\
\hdashline Boryor et al. (2009) \cite{boryor2009downloadable} & ~1$^{\star}$ & Single Tooth & $\mu$CT & EDP, Adaptive & H, Uniform & CC, Adaptive & CG & Non-CM & Yes & Yes & No \\
OpenMandible (2021) \cite{vukicevic2021openmandible} & ~1$^{\star}$ & 1 Mandible & CBCT + $\mu$CT & EDP & H, Uniform & O, Adaptive & CG & Non-CM & Yes & Yes & Yes \\
OpenJaw (2021) \cite{OpenJawDataset,gholamalizadeh2021multi} & 3 & 3 Mandibles & CBCT & H, Uniform & H, Uniform & H, Adaptive & Non-CG & Non-CM & Yes & No & No \\
Open-Full-Jaw (our study) & ~17 {$^{*}$} & 29 Jaws {$^{*}$} & CBCT & H, Adaptive & H, Uniform & H, Adaptive & CG & CM & Yes & Yes & Yes \\ \hline
\end{tabular}%
\vspace{1mm}
\begin{tablenotes}
\small \textbf{N/A}: Not Applicable; \textbf{DM}: Dental Model; \textbf{H}: Homogeneous; \textbf{EDP}: Enamel, Dentine, and Pulp; \textbf{CC}: Cortical and Cancellous bone; \textbf{O}: Orthotropic bone, further details can be found in \cite{vukicevic2021openmandible};\\
\textbf{CG}: Congruent interface (no gap/penetration in the contacting interfaces); \textbf{Non-CG}: Non-congruent interface; \textbf{CM}: Conformal Mesh; \textbf{Non-CM}: Non-Cnformal Mesh.\\
\item{$\star$ Obtained from cadaver or dried skull. * See Table \ref{table:scan_info} for further details., $\dagger$ Non-smooth top surface of PDL, see \cite{kawamura2019biomechanical,kawamura2022biomechanical,sarrafpour2013tooth} for illustrations.}
\end{tablenotes}
\end{threeparttable}
}
\label{table:lit_review}
\end{table*}

The main reason for using a single model in the literature is that developing accurate computational models of the human jaw is challenging and involves time-consuming and labor-intensive processes such as segmentation, geometry reconstruction, geometry processing, re-meshing, and mesh simplification tasks. For instance, generating a complex and highly detailed finite element (FE) model of the entire human jaw can take up to several months per scan \cite{vukicevic2021openmandible}. Therefore, developing several patient-specific FE models may not be feasible for many researchers. In addition, currently, there are no publicly available datasets of full dentition human jaw to be used by researchers, except for two studies \cite{vukicevic2021openmandible,OpenJawDataset} with a limited number of studied subjects.

In other words, in almost all of the studies focusing on full dentition or single/multiple tooth analysis \cite{benaissa2020stress,boryor2009downloadable,seo2021comparative,ding2014influence,savignano2020three,ortun2020silico,benaissa2020stress,oenning2018resorptive,sarrafpour2013tooth,lee2018biomechanical}, the utilized geometries, volumetric meshes, and FE models have not been made publicly available, which makes it difficult to reproduce and compare the results. Table \ref{table:lit_review} presents an overview of the related studies in the literature by providing details on the studied cohort, discretization type, and availability of the models.

As one of the few studies with public data, the OpenMandible \cite{vukicevic2021openmandible} provides detailed geometries of one mandible structure and all teeth obtained from a dried male skull. The study scans the mandible in two different steps to provide detailed teeth structures, i.e., pulp and enamel. First, the mandible was scanned using a CBCT scanner with a voxel size of 0.133 mm. Second, all mandibular teeth were removed from the bone before being scanned by the micro-computed tomography (micro-CT or $\mu$CT). Micro-CTs are high-resolution CT scans that are normally acquired from dead specimens or cadavers due to high x-ray exposures. However, the detailed geometries obtained from a single mandible cannot cover geometrical variations across different patients in the population.

The recently introduced OpenJaw Dataset \cite{OpenJawDataset} provides open-access reconstructed geometries of three patients' mandibles acquired from CBCT scans. Each patient’s data includes surface meshes of the reconstructed mandible, teeth, and PDL geometries. In the geometry reconstruction step, the utilized scans in the study (one with 0.15 mm and two with 0.3 mm voxel sizes) were upsampled to the same resolution of 0.15 mm \cite{gholamalizadeh2021multi}. Although this study provides more public samples, generalizability to the population remains a problem. Moreover, both of the abovementioned studies include manual tasks in different geometry-processing or meshing tools making it difficult for other researchers to reproduce their meshes using the unprocessed reconstructed meshes.

Our main contributions in the Open-Full-Jaw study can be summarized as, 
\begin{enumerate}
   \item We provide an open-access dataset of different patient-specific models of the human jaw, including the maxilla, mandible, full dentition, and the PDL geometries obtained from CBCT scans of 17 patients. It is the largest publicly available dataset with validated segmented geometries and quality volumetric meshes that can directly be used in FEM studies. Furthermore, to the best of our knowledge, it is the first repository containing the maxillary jaws of different patients.
   \item We introduce a unique repository (\url{https://github.com/diku-dk/Open-Full-Jaw}) containing (1) clinically validated segmented geometries and the resulting dense irregular surface meshes; (2) the quality and adaptive volumetric/surface meshes to be used in FE simulations; (3) the automatically generated FEM files for tipping and biting scenarios used for the FE analysis of this work; (4) the principal axes of every patient's tooth providing great information for the users to automatically set up different loading conditions.
  \item For reproducibility, we share our pipeline developed based on open-source meshing tools \cite{ftetwild,libigl} to generate the models of this study. This python-based library automates the FE model generation process, including geometry processing and re-meshing tasks with minimal human intervention, by setting a few required parameters.
   \item This pipeline allows other researchers in the field to generate quality volumetric meshes and FE models directly from dense and uncleaned meshes with minimal human intervention. This will help other researchers to easily extend their datasets without spending much time and effort on manually cleaning up the meshes and non-trivially producing conformal meshes. 
   \item Our pipeline ensures conformal meshes in the contacting interfaces without any undesired gaps or penetrations and provides adaptive meshes that are vital for reducing the total number of elements while using finer meshes in specific regions, e.g., teeth sockets, alveolar crest, and alveolar process.
\end{enumerate}
   
All in all, we believe the Open-Full-Jaw dataset can be used for various intra- and inter-patient analyses \cite{gholamalizadeh2020mandibular,savignano2020three} such as intact tooth movement modeling, bite force estimation, restorative procedures modeling including cavity fillings and dental implants, just to name a few. Besides, it can greatly impact the reproducibility of future studies. For the reproducibility of this study, we use open-source meshing tools. Still, the reconstructed geometries provided in our dataset can be imported into any desired open-source or commercial meshing tools or FE frameworks to re-mesh and generate computational models.\\

\section{Important traits required for a successful FEM}

Developing a patient-specific FE model begins with segmenting/annotating the desired regions of the medical scan obtained from the patients. Next, the segmented regions are reconstructed as surface meshes generally composed of irregular dense meshes with no guarantees of manifoldness, watertightness, or absence of self-intersection, which are crucial for developing stable and accurate computational models. Hence, one needs to generate quality meshes from the exported dense and irregular meshes that are not necessarily guaranteed to have the mentioned criteria.

Moreover, different preprocessing steps such as geometry processing, mesh reduction, and re-meshing are essential for developing FE models from image-based reconstructed geometries. When modeling geometries with shared contacting interfaces, each of the mentioned processes can produce errors on the contacting surfaces and result in undesired gaps/penetrations between them. In the cases where two adjacent segments are watertight, it is still challenging to discretize the segmented domains such that they agree on the same discretization on the shared contacting interfaces. Therefore, the focus of this section is on essential aspects needed to be considered for the discretization of the computational domains with shared contacting interfaces; we also discuss potential options for developing a proper FE model of the human jaw.

\begin{figure*}[t]
\centering
    \includegraphics[width=\textwidth]{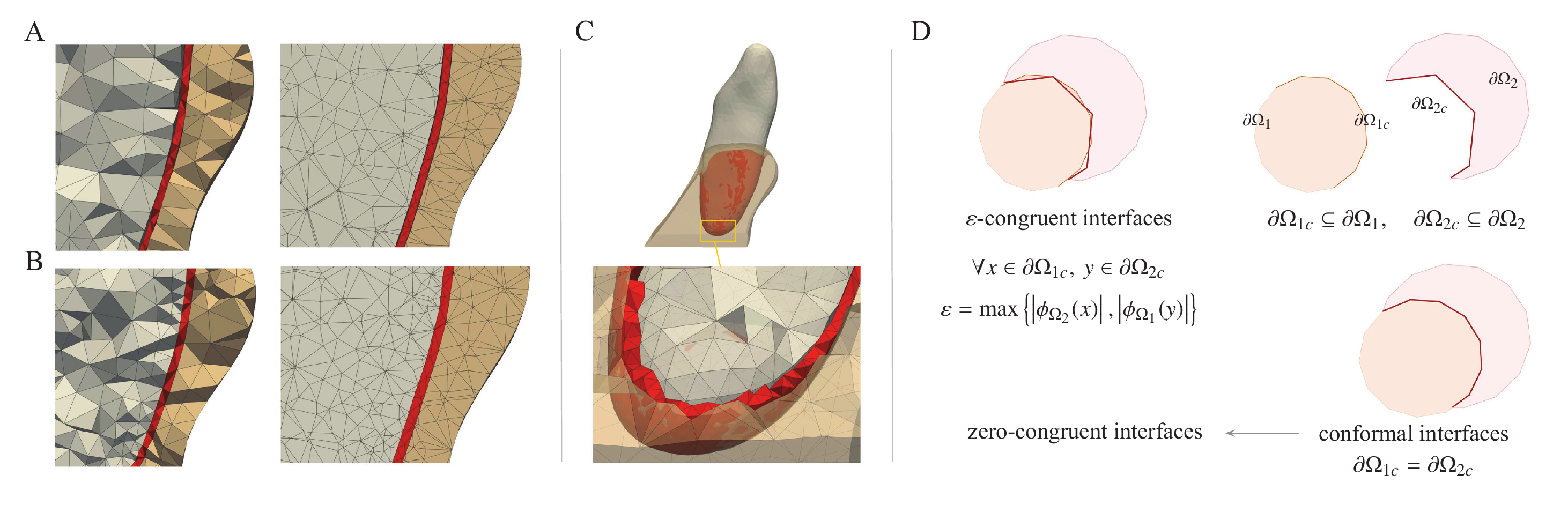}
    \vspace{-8mm}
    \caption{A visualized comparison between the $\varepsilon$-congruent and conformal meshes. \textbf{A}: the $\varepsilon$-congruent contacting surfaces with gaps/penetrations in the contacting interfaces. \textbf{B}: the same structure with the multi-domain discretization resulting in conformal meshes. \textbf{C}: a close-up cross-section view of the gaps/penetrations at the root level. \textbf{D}: a schematic illustration of the $\varepsilon$-congruent and conformal meshes.}
    \label{fig:gaps}
\end{figure*}

\subsection{Congruent contacting interfaces}

The discretization of the computational domain is an essential step for developing computational models considering the biomechanical behavior of tissues having sliding or bilateral contact with other domains/tissues. It mainly affects the numerical stability of the computational models when soft structures, e.g., PDL, contact hard tissues like bone or teeth. Besides, a coarse or poor-quality discretization can itself cause a locking effect and influence the accuracy of the stress/strain concentrations compared to that of the values measured in vivo.

When two contacting domains are reconstructed, discretized, or re-meshed independently, undesired gaps/penetrations are inevitable, causing two different boundary definitions for an identical contacting interface between the two adjacent/contacting domains. The agreement of the two contacting geometries on the contacting boundary or shared interface can be analyzed in two different levels; first, on the geometry level and based on the curvature of the contacting surfaces; second, on the mesh-based level focusing on the agreement on the identical discretization of the contacting interface in terms of the position of vertices, edges, and faces. The former is called here the ``interface congruency". In the biomechanical field of study, different theoretical and computational studies analyze the effect of the ``ball-and-socket" joint congruencies such as in shoulder, hip, and temporomandibular joints to analyze their instabilities and dislocations under different circumstances \cite{ernstbrunner2016biomechanical,beek2000three}. It should be noted that the current study only focuses on developing computational models of human mandible and maxilla for tooth movements and the mesh congruency of the contacting surfaces instead of congruency of the ``ball-and-socket" joints, and the utilized ``congruency" term needs to be distinguished from those in biomechanical studies for analyzing the joint congruencies \cite{ernstbrunner2016biomechanical,beek2000three}.

As mentioned before, by using the conventional one-by-one domain discretization, mesh reduction, or quality meshing processes, it is challenging to achieve congruent surfaces due to the generation of gaps/penetrations between two contacting surfaces (see Figure \ref{fig:gaps}). We propose using signed distance fields in the contacting surfaces to quantitatively evaluate the error between two contacting regions. The signed distance function, $\phi_{\Omega}(x)$, for a domain $\Omega \subset R^3$ and an arbitrary point $x \in R^3$ is defined as
\begin{equation} \label{eq:sdf}
 \phi_{\Omega} (x)=\begin{cases}
    -d(x, \partial \Omega), & \text{if $x$ }\in \Omega,\\
    0, & \text{if $x$ }\in \partial \Omega,\\
    d(x, \partial \Omega), & \text{if $x$ }\in \Omega^c,\\
  \end{cases}
\end{equation}
\noindent where $\partial \Omega$ denotes the boundary of the domain $\Omega$, $\Omega^c$ represents the complement of $\Omega$, and $d(x, \partial \Omega)$ indicates the Euclidean distance between the point $x$ and the boundary of the domain $\partial \Omega$. We measure the congruency of the surfaces by calculating the signed distance field of each contacting surface with respect to the other domain as
\begin{equation} \label{eq:eps_cong}
\begin{split}
\varepsilon = \max \left\{  \left|  \phi_{\Omega_{2}} (x)\right|, \left|  \phi_{\Omega_{1}} (y)\right| \right\}, \\
\forall x  \in \partial \Omega_{1c}, \: y \in \partial \Omega_{2c}, \\
\partial \Omega_{1c} \subseteq \partial \Omega_{1}, \; \partial \Omega_{2c} \subseteq \partial \Omega_{2},
\end{split}
\end{equation}
where $\partial \Omega_{1}$ and $\partial \Omega_{2}$ denote the boundaries of the domains $\Omega_{1}$ and $\Omega_{2}$, respectively, and $\partial \Omega_{1c}$ and $ \partial \Omega_{2c} $ refer to the contacting surfaces of the domains. The computed $\varepsilon$ values for two contacting domains are then called epsilon-congruent surfaces. The epsilon value indicates the measurement error, where the values close to zero refer to completely congruent surfaces.

\begin{figure*}[t]
	\centering
	\vspace{-5mm}
	\includegraphics[width=\textwidth]{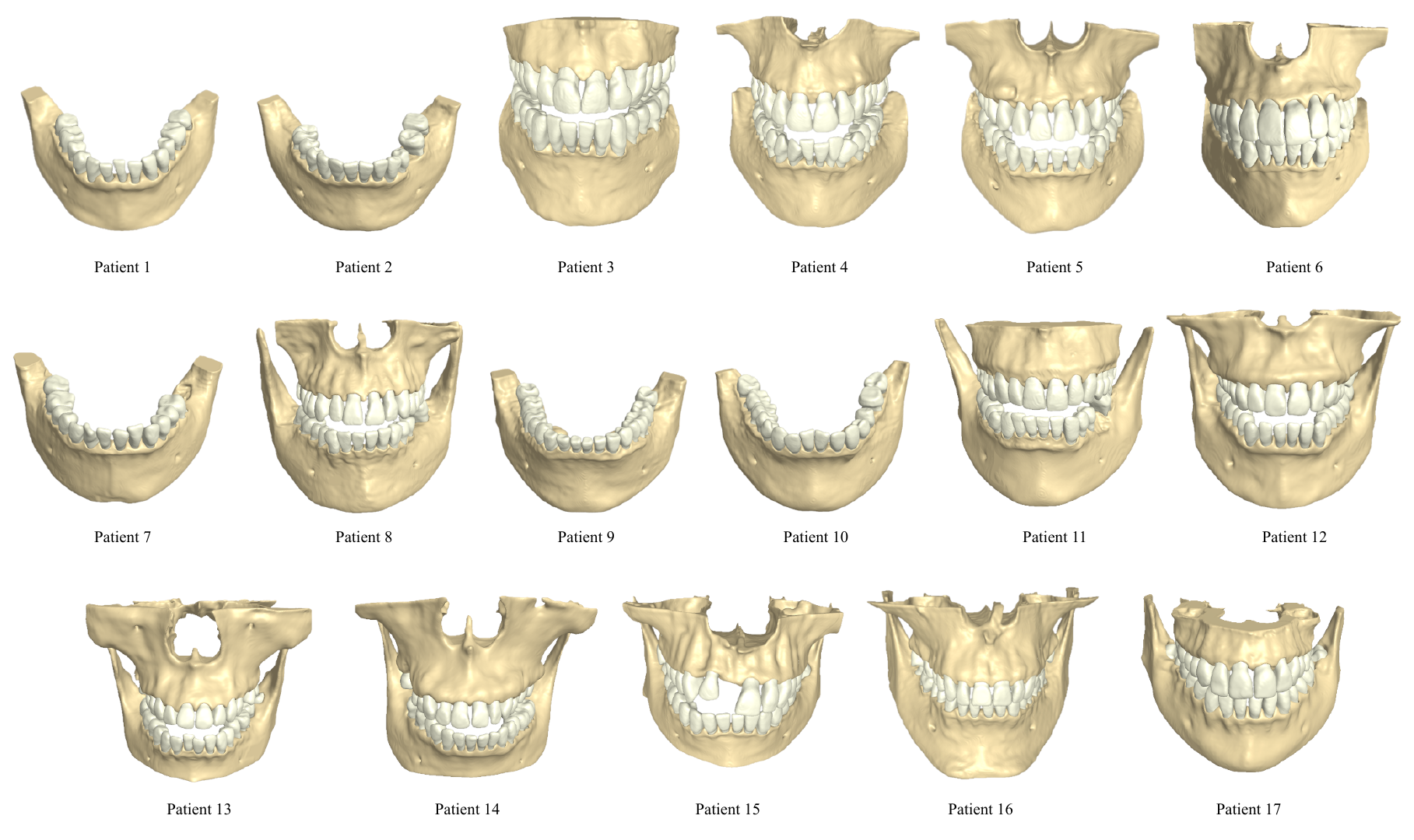}
 	\vspace{-8mm}
 	\caption{Reconstructed 3D geometries from 17 CBCT scans. Note that the maxillary jaw is removed in some scans due to having several metal artifacts and/or missing teeth. The anatomical variations in different jaws and teeth of our dataset indicate the necessity for introducing such a model repository for assessing the generalizability of the FEM results in the related clinical/population studies.}
 	\label{fig:gallery}
\end{figure*}

\subsection{Conformal mesh interfaces}

A special case for the congruent surfaces is ``conformal meshes'' or ``mesh conformity'' and can be described as the identical discretization of the contacting interfaces. That is to say, for two contacting domains/geometries like $\Omega_{1}$ and $\Omega_{2}$, the contacting surfaces $\partial \Omega_{1c}$ and $ \partial \Omega_{2c} $ are assumed identical. More specifically, they share identical vertices, edges, and elements on the contacting interfaces. In general, applying the conformal mesh criterion on multi-domains with contacting surfaces is a challenging process, as most of the geometry and mesh processing algorithms used in different free software produce the quality meshes per domain independently. Therefore, the contacting surfaces should be identified and combined as a different step to create identical contacting interfaces on the mesh of each domain. 

Finally, the volumetric meshes can be generated by enforcing the meshing algorithm to fully preserve the input surface mesh. This raises questions about the final mesh quality, as the additional surface preservation constraint affects the mesh generation and the optimization algorithms. Hence, the best approach for generating conformal meshes is to consider all contacting domains simultaneously while generating the volumetric meshes.

Generating multi-domain volumetric meshes would generate zero-congruent and conformal meshes that omit any numerical errors caused by non-congruent contacting meshes. Therefore, it is essential to utilize a meshing algorithm that generates volumetric meshes considering multi-domain boundaries and performing boolean operations on the input surface meshes by using an implicit representation of the input meshes. In this study, we use fTetWild \cite{ftetwild} that supports all aspects mentioned above for multi-domain volumetric meshes. This provides a volumetric mesh of teeth-PDL-bone geometries in which there are shared points in the teeth-PDL and PDL-bone contacting surfaces.

\subsection{Proper modeling of the PDL layer}
PDL is a thin structure that connects alveolar bone to the teeth cementum and acts as a shock absorber in the chewing or biting process. It also plays an important role in transferring load from teeth to the bone in orthodontic treatments; when triggered with enough orthodontic forces, it results in a bone remodeling process. As this thin and soft structure shares interfaces with two hard tissues of the teeth and bone geometries, the computational model may become unstable if the PDL geometry does not have congruent surfaces with teeth and bone in the contacting interfaces. \label{traits:pdl}

\begin{table*}[t]
\caption{Specifications of the utilized CBCT scans. Different scans with various filed-of-views and slice thicknesses are used in this study. All scans are converted to an identical slice thickness of 0.15 mm to avoid biases in the geometry reconstruction process, especially when applying smoothing filters to eliminate noises on the segmented teeth and bone geometries. Note that the maxillary jaws of the patients with many metal artifacts and/or missing teeth are removed.}
\vspace{1.5mm}
\resizebox{\textwidth}{!}{%
\begin{tabular}{lcccccccc}
\hline
\multicolumn{1}{c}{\multirow{2}{*}{\begin{tabular}[c]{@{}c@{}}Patient's~~\\ ID\end{tabular}}} & \multicolumn{2}{c}{Original scan} & \multicolumn{2}{c}{Final resampled ROI} & \multicolumn{2}{c}{Jaws} & \multicolumn{2}{c}{Periodontal disease}\\ \cline{2-9} 
\multicolumn{1}{c}{} & Dimension & Slice thickness (mm)& Dimension & Slice thickness (mm) & Mandible & Maxilla & Mandible & Maxilla\\ \hline
\rowcolor{Gray}
Patient 1  & ~~~$400\times400\times280$ & 0.3  &  ~$676\times530\times280$ & 0.15 & Yes &  -  & No & Missing teeth\\
Patient 2  & ~~~$400\times400\times280$ & 0.3  &  ~$670\times440\times344$ & 0.15 & Yes &  -  & No & Missing teeth\\
\rowcolor{Gray}
Patient 3  & ~~~$532\times532\times540$ & 0.15 &  ~$534\times435\times338$ & 0.15 & Yes & Yes & Moderate & Moderate \\
Patient 4  & ~~~$532\times532\times540$ & 0.15 &  ~$534\times435\times338$ & 0.15 & Yes & Yes & Mild & Mild bone loss \\
\rowcolor{Gray}
Patient 5  & ~~~$532\times532\times540$ & 0.15 &  ~$525\times425\times290$ & 0.15 & Yes & Yes & No & No\\
Patient 6  & ~~~$534\times534\times430$ & 0.15 &  ~$534\times534\times430$ & 0.15 & Yes & Yes & No & No\\
\rowcolor{Gray}
Patient 7  & ~~~$400\times400\times280$ & 0.3  &  ~$614\times470\times270$ & 0.15 & Yes &  -  & No & - \\
Patient 8  & ~~~$400\times400\times280$ & 0.3  &  ~$800\times800\times560$ & 0.15 & Yes & Yes & Moderate & Moderate\\
\rowcolor{Gray}
Patient 9  & ~~~$400\times400\times280$ & 0.3  &  ~$525\times425\times290$ & 0.15 & Yes &  -  & No & -\\
Patient 10 & ~~~$400\times400\times280$ & 0.3  &  ~$525\times425\times290$ & 0.15 & Yes &  -  & No & -\\
\rowcolor{Gray}
Patient 11 & ~~~$400\times400\times280$ & 0.3  &  ~$525\times425\times290$ & 0.15 & Yes & Yes & No & No\\
Patient 12 & ~~~$400\times400\times280$ & 0.3  &  ~$525\times425\times290$ & 0.15 & Yes & Yes & Moderate & Moderate\\
\rowcolor{Gray}
Patient 13 & ~~~$400\times400\times280$ & 0.3  &  ~$525\times425\times290$ & 0.15 & Yes & Yes & ~No$^\dagger$ & ~No{$^{*}$}\\
Patient 14 & ~~~$400\times400\times280$ & 0.3  &  ~$525\times425\times290$ & 0.15 & Yes & Yes & No & No\\
\rowcolor{Gray}
Patient 15 & ~~~$750\times750\times400$ & 0.2  &  ~$525\times425\times290$ & 0.15 & Yes & Yes & ~No$^\dagger$ & ~No{$^{*}$}\\
Patient 16 $^\star$ & ~~~$520\times406\times340$ & 0.25 &  ~$866\times636\times566$ & 0.15 & Yes & Yes & No & No\\
\rowcolor{Gray}
Patient 17 & ~~~$500\times500\times500$ & 0.2  &  ~$668\times530\times364$ & 0.15 & Yes & Yes & ~No$^\dagger$ & No\\
\hline
\end{tabular}%
}
\begin{tablenotes}
\footnotesize $\star$ The scan obtained from 3DSlicer's ``Sample Data'' module, titled ``CBCT-MRI Head''.\\
$\dagger$ The mandible includes partially erupted wisdom tooth/teeth.\\
* The maxilla includes an impacted wisdom tooth.
\end{tablenotes}
\label{table:scan_info}
\end{table*}

\subsection{Quality mesh generation}

Tetrahedralization algorithms usually require a set of closed, self-intersection free surface meshes as input, which is a requirement that is challenging to guarantee in our setting without manual interaction. In fact, traditional medical imaging pipeline require the large majority of the manual processing in this cleanup phase, as self-intersection, holes, or other imperfections normally appear when processing real-world data, and are especially common (and time consuming) when dealing with thin layers, like the PDL layer. 

We propose a very different approach, where we accept that these imperfection exist, and design a meshing pipeline that tolerates and automatically heals them. We base our approach on the method recently introduced in \cite{hu2018tetrahedral,hu2020fast}: instead of meshing one domain at a time, TetWild meshes the entire volume of the bounding box containing the soup of triangles of all surfaces of interest. The triangles are approximated with faces of the tetrahedral mesh. This procedure does not require clean input geometry, it can tolerate degenerate input triangles, self-intersections, and holes \cite{hu2018tetrahedral}. In the original TetWild algorithms, the final mesh is extracted using a robust filtering procedure based on either flood fill or the generalized winding number \cite{Jacobson-13-winding}, in both cases assuming that the user is interested in a single material mesh. We propose a novel filtering procedure for the filtering of multi-material tetrahedral meshes common in medical imaging in Section \ref{section:tet_filtering}, and we show that our extension of TetWild is ideal for constructing our dataset, as it removes the expensive and tedious manual cleanup of the input surface meshes.

\section{Controlling shared interfaces using volume mesh generation}

This section reviews the entire process for developing the FE models of 17 jaws presented in Figure \ref{fig:gallery}. To be more specific, Section \ref{section:criteria} describes the utilized criteria for the cohort selection process; Sections \ref{section:segmentation} and \ref{section:clinical_val} provide a detailed description for the CBCT segmentation and clinical validation; the geometrical variations of the reconstructed mandibles and maxillae are investigated in Section \ref{section:geom} based on widely used clinical landmarks; and finally, the different steps of the used pipeline for developing high-quality volumetric meshes of human jaws are studied in Section \ref{section:auto_pipe}.

\subsection{Cohort selection}
\label{section:criteria}
We use CBCT scans of 17 different patients with different voxel resolutions from 3Shape A/S in-house dataset. Various criteria including the original voxel size of the scan, minimal metal filling artifacts, and the absence of implants and severe periodontal diseases are considered in selecting the mentioned cohort. Also, the selected cohort has no evidence of maxillofacial surgery or skeletal diseases. Sensitive information of the patients such as name, age, and gender are stripped due to the General Data Protection Regulation (GDPR) rules. The utilized scans were acquired from the patients by their associated doctors/orthodontists as a part of treatment plans, and the authors had no roles in the acquisition process.

\begin{figure*}[t]
\centering
\includegraphics[width=0.95\textwidth]{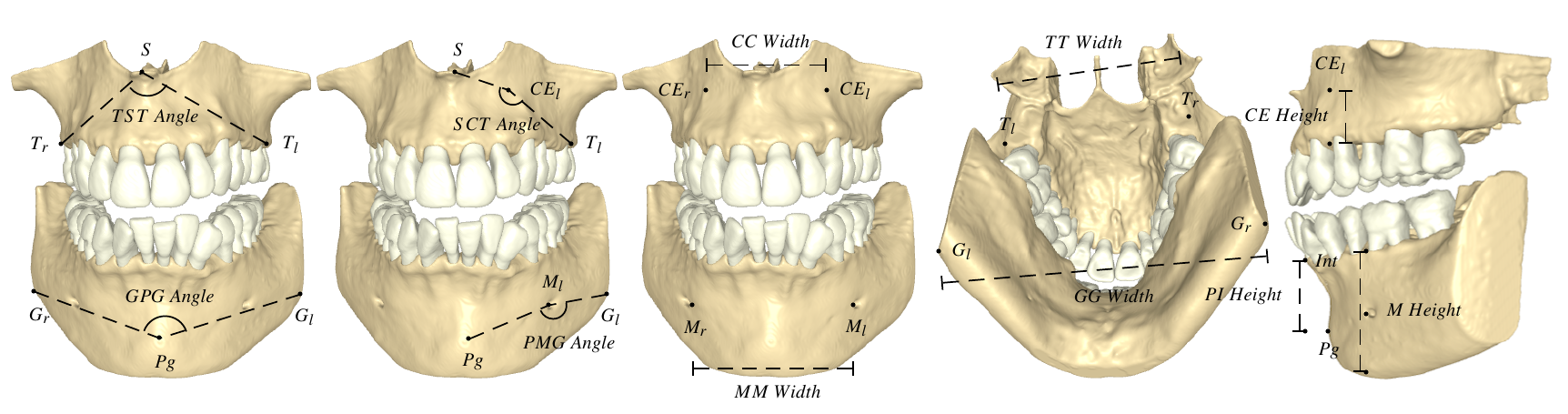} 
\caption{The utilized morphological landmarks for analyzing and quantification of the geometrical variation of the reconstructed mandibles and maxillae in different models.}
\label{fig:landmarks}
\end{figure*}

\begin{table*}[]
\centering
\caption{Description of the utilized morphological landmarks obtained based on the literature \cite{buikstra1994standards,tunis2017sex,bayome2013new,vallabh2020morphology, villanueva2017gender}, as illustrated in Figure \ref{fig:landmarks}.}
\vspace{1.5mm}
\label{table:landmarks_def}
\resizebox{\textwidth}{!}{%
\begin{tabular}{lll}
\hline
Jaw type & Landmark & Landmark's description \\ \hline
\multicolumn{1}{l}{\multirow{6}{*}{Mandible}} & MM Width & Maximum breadth of mandibular body at the mental foramen  \cite{buikstra1994standards} \\
\multicolumn{1}{l}{} & MF Height & Body height: the height of the mandible at the mental foramen ($M_{l}$) \cite{tunis2017sex}\\
\multicolumn{1}{l}{} & PMG Angle & Mandibular body angle: the angle between the lines connecting the left mental foramen ($M_{l}$) to pogonion ($Pg$) and left gonion, inspired from \cite{bayome2013new}  \\
\multicolumn{1}{l}{} & GPG Angle & Pogonion angle: the angle between the lines connecting pogonion ($Pg$) to the left and right gonia ($G_{l}, G_{r}$), inspired from \cite{buikstra1994standards,villanueva2017gender} \\
\multicolumn{1}{l}{} & GG Width & Intergonial distance: direct distance between the left and right gonia \cite{vallabh2020morphology}\\
\multicolumn{1}{l}{} & PI Height & The perpendicular distance between the anterior interdental ($Int$) and pogonion ($Pg$) \cite{vallabh2020morphology} \\ \hline
\multicolumn{1}{l}{\multirow{5}{*}{Maxilla}} & CC Width & Anterior maxillary basal width: the length of the line connecting the left and right canine eminences ($CE_{l}, CE_{r}$) \cite{bayome2013new} \\
\multicolumn{1}{l}{} & CE Height & The perpendicular distance from the left canine eminence ($CE_{l}$) to the alveolar crest\\
\multicolumn{1}{l}{} & SCT Angle & The angle between the lines connecting the left canine eminence ($CE_{l}$) to the anterior nasal spine ($S$) and the left maxillary tuberosity ($T_{l}$)  \\
\multicolumn{1}{l}{} & TST Angle & Anterior nasal spine angle: the angle between the lines connecting the anterior nasal spine ($S$) to the left and right maxillary tuberosity ($T_{l}, T_{r}$) \\
\multicolumn{1}{l}{} & TT Width & Posterior maxillary width; the distance between the left and right maxillary tuberosity ($T_{l}, T_{r}$), inspired from \cite{bayome2013new} \\ \hline
\end{tabular}%
}
\end{table*}

\subsection{Data specifications and geometry reconstruction}
\label{section:segmentation}
To reconstruct the patient-specific geometries, first, the scans are imported in 3DSlicer \cite{fedorov20123d} in the standard Digital Imaging and Communications in Medicine (DICOM) format. Table \ref{table:scan_info} provides details of the scans utilized in this study. Next, according to the pre-evaluation criteria (metal fillings or implant artifacts), we decide on which jaws are suitable to be segmented from the scan.

The resolution of the selected CBCT scans is at most 0.3 mm since based on our experience, accurate tooth-bone segmentation of the scans with slice thicknesses above 0.3 mm is very challenging. Besides, applying smoothing functions with identical kernel sizes results in smoother segments for coarse voxel sizes, making it difficult to remove the segmentation noise (e.g., rugged surfaces) while preserving the fine details in desired regions (e.g., the alveolar crests of teeth sockets and root apexes). As a result, to avoid biases in the geometry reconstruction process, we convert all scans to an identical resolution of 0.15 mm, as summarized in Table \ref{table:scan_info}, by oversampling a cropped region of interest (ROI) containing the jaws using linear interpolation.

The tooth-bone segmentation is performed based on the semi-automatic watershed algorithm \cite{soille2013morphological} provided in 3DSlicer's SegmentEditorExtraEffects extension. The result of the watershed algorithm is then refined to correct the misclassified teeth and bone. Later, the geometries are smoothened using the 3DSlicer's standard \textit{median} and \textit{joint smoothing} modules \cite{pinter2019polymorph,taubin1996optimal}. The joint smoothing method \cite{taubin1996optimal} smooths the adjacent segments simultaneously and enforces watertight interfaces between them. The tooth-bone segmentation procedure proceeds until the segmentation accuracy meets our clinical validation criteria (see Section \ref{section:clinical_val}). Finally, the segmented regions are exported as surface meshes in the Object file format (OBJ).

\subsection{Clinical validation of the teeth-bone segmentation}\label{section:clinical_val}
The segmented teeth-bone geometries of all patients are validated by clinical experts. This is done by identifying the existence of any periodontal diseases and categorizing each patient into one of the no, mild, or moderate periodontal disease categories. The patients with severe periodontal diseases are excluded from the analyses based on the cohort selection criteria mentioned in Section \ref{section:criteria}. Moreover, we assess the accuracy of segmentation in areas close to teeth/bone borders, cervical regions of bone around teeth sockets, tooth-bone interfaces, and roots.

\subsection{Geometrical variations of the dataset}
\label{section:geom}

We evaluate the geometrical variation of the reconstructed geometries using 3D morphological and cephalometric landmarks and measurements adapted or inspired from the literature \cite{buikstra1994standards,tunis2017sex,bayome2013new,vallabh2020morphology, villanueva2017gender}, as illustrated in Figure \ref{fig:landmarks} and described in Table \ref{table:landmarks_def}. Accordingly, the measured values per patient and the mean and standard deviation of each morphological measurement across all patients are listed in Table \ref{table:landmarks}. Note that even though the standard deviation of each measurement may seem rather small, such small changes can lead to significant variations in the overall shape of the jaw, indicating the availability of high geometrical variations among different jaws under the study.

\begin{table*}[t]
\caption{Assessment of the geometrical variations of the mandibles and maxillae based on the computed measurements illustrated in Figure \ref{fig:landmarks}. Note that even small changes in the values can change the overall shape of the jaw considerably, indicating that high geometrical variations are available within our models.}
\label{table:landmarks}
\vspace{1.5mm}
\resizebox{\textwidth}{!}{%
\begin{tabular}{lccccccccccc}
\hline
\multicolumn{1}{c}{\multirow{2}{*}{\multirow{2}{*}{\begin{tabular}[c]{@{}c@{}}Patient's~~\\ ID\end{tabular}}}} & \multicolumn{6}{c}{Mandibular jaw} & \multicolumn{5}{c}{Maxillary jaw} \\ \cline{2-12} 
\multicolumn{1}{c}{} & \begin{tabular}[c]{@{}c@{}}MM Width\\ (mm)\end{tabular} & \begin{tabular}[c]{@{}c@{}}MF Height\\ (mm)\end{tabular} & \begin{tabular}[c]{@{}c@{}}PMG Angle\\ (deg)\end{tabular} & \begin{tabular}[c]{@{}c@{}}GPG Angle\\ (deg)\end{tabular} & \begin{tabular}[c]{@{}c@{}}GG Width\\ (mm)\end{tabular} & \begin{tabular}[c]{@{}c@{}}PI Height\\ (mm)\end{tabular} & \begin{tabular}[c]{@{}c@{}}CC Width\\ (mm)\end{tabular} & \begin{tabular}[c]{@{}c@{}}CE Height\\ (mm)\end{tabular} & \begin{tabular}[c]{@{}c@{}}SCT Angle\\ (deg)\end{tabular} & \begin{tabular}[c]{@{}c@{}}TST Angle\\ (deg)\end{tabular} & \begin{tabular}[c]{@{}c@{}}TT Width\\ (mm)\end{tabular} \\ \hline
\rowcolor{Gray}
Patient 1 & 49.30 & 28.94 & 147.20 & 77.50 & 95.32 & 21.70 & - & - & - & - & - \\
Patient 2 & 46.54 & 27.31 & 142.30 & 79.00 & 83.54 & 21.45 & - & - & - & - & - \\
\rowcolor{Gray}
Patient 3 & 46.89 & 33.31 & 131.70 & 86.60 & 69.80 & 26.43 & 40.84 & 15.62 & 141.60 & 68.40 & 63.66 \\
Patient 4 & 41.61 & 32.73 & 141.40 & 81.20 & 73.02 & 19.94 & 30.21 & 12.24 & 146.90 & 61.10 & 52.62 \\
\rowcolor{Gray}
Patient 5 & 42.51 & 25.57 & 158.70 & 96.50 & 78.20 & 16.41 & 29.92 & 12.88 & 154.50 & 67.80 & 59.87 \\
Patient 6 & 38.42 & 26.72 & 151.80 & 85.80 & 71.61 & 19.71 & 30.09 & 16.09 & 143.50 & 66.70 & 55.36 \\
\rowcolor{Gray}
Patient 7 & 47.18 & 27.39 & 140.20 & 71.90 & 82.10 & 21.07 & - & - & - & - & - \\
Patient 8 & 41.19 & 29.48 & 146.10 & 75.30 & 88.88 & 21.63 & 27.68 & 13.04 & 148.60 & 68.30 & 54.78 \\
\rowcolor{Gray}
Patient 9 & 50.31 & 32.33 & 142.40 & 81.70 & 94.29 & 26.53 & - & - & - & - & - \\
Patient 10 & 48.34 & 33.14 & 150.80 & 77.60 & 86.19 & 23.02 & - & - & - & - & - \\
\rowcolor{Gray}
Patient 11 & 51.29 & 32.36 & 151.40 & 80.50 & 97.08 & 27.21 & 37.13 & 17.08 & 129.20 & 69.90 & 61.01 \\
Patient 12 & 46.34 & 33.96 & 154.00 & 81.30 & 97.10 & 26.21 & 35.83 & 14.70 & 133.20 & 69.90 & 68.12 \\
\rowcolor{Gray}
Patient 13 & 45.62 & 19.24 & 147.00 & 77.90 & 89.11 & 17.86 & 33.46 & 12.49 & 132.40 & 80.30 & 61.51 \\
Patient 14 & 49.39 & 23.02 & 145.00 & 91.90 & 89.42 & 17.68 & 34.65 & 13.30 & 147.60 & 77.30 & 72.85 \\
\rowcolor{Gray}
Patient 15 & 45.15 & 20.01 & 147.80 & 75.80 & 91.21 & 18.83 & 25.90 & 17.43 & 131.70 & 63.70 & 58.99 \\
Patient 16 & 46.76 & 36.57 & 151.40 & 71.70 & 104.70 & 26.88 & 35.94 & 17.00 & 140.20 & 74.60 & 71.66 \\
\rowcolor{Gray}
Patient 17 & 40.21 & 24.76 & 148.20 & 70.20 & 91.76 & 19.16 & 27.61 & 11.70 & 137.50 & 67.70 & 53.33 \\ \hline
\begin{tabular}[c]{@{}l@{}}Mean\\ $\pm$ STD\end{tabular} & \begin{tabular}[c]{@{}c@{}}45.71\\ $\pm$ 3.73\end{tabular} & \begin{tabular}[c]{@{}c@{}}28.64\\ $\pm$ 5.03\end{tabular} & \begin{tabular}[c]{@{}c@{}}146.91\\ $\pm$ 6.24\end{tabular} & \begin{tabular}[c]{@{}c@{}}80.14\\ $\pm$ 7.00\end{tabular} & \begin{tabular}[c]{@{}c@{}}87.25\\ $\pm$ 9.79\end{tabular} & \begin{tabular}[c]{@{}c@{}}21.87\\ $\pm$ 3.58\end{tabular} & \begin{tabular}[c]{@{}c@{}}32.44\\ $\pm$ 4.55\end{tabular} & \begin{tabular}[c]{@{}c@{}}14.46\\ $\pm$ 2.10\end{tabular} & \begin{tabular}[c]{@{}c@{}}140.58\\ $\pm$ 7.96\end{tabular} & \begin{tabular}[c]{@{}c@{}}69.64\\ $\pm$ 5.43\end{tabular} & \begin{tabular}[c]{@{}c@{}}61.15\\ $\pm$ 6.85\end{tabular} \\ \hline
\end{tabular}%
}
\end{table*}

\subsection{The pipeline for generating FE model of a human jaw}
\label{section:auto_pipe}

The flexibility of the FE method allows it to use a wide range of spatial discretizations \cite{femtable}. We opt for an unstructured tetrahedral mesh as can be robustly generated using automatic meshing tools \cite{ftetwild} and can lead to similar accuracy and running time when using high order elements as structured meshes \cite{schneider2019large}.

\begin{figure*}[t]
\centering
    \includegraphics[width=\textwidth]{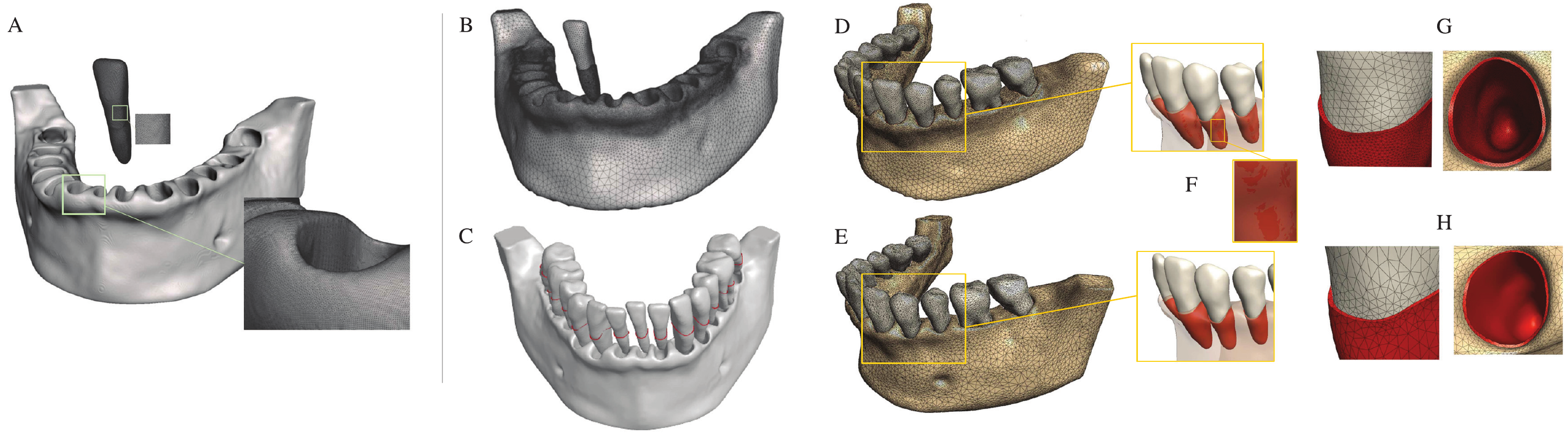}
    \caption{Finite element models created based on irregular dense meshes (\textbf{left subfigure, A}) exported from the segmentation step. \textbf{Right subfigure:} A comparison between the results of the conventional meshing approach (\textbf{top-row)} and utilized pipeline \textbf{(bottom-row)}. The conventional approach involves time-consuming and labor-intensive geometry and mesh processing tasks \textbf{(B)}; this results in non-congruent contacting interfaces \textbf{(F)}, and non-conformal meshes \textbf{(G)}. Note that the spotted marks in \textbf{F} indicate the undesired gaps/penetrations in the contacting interfaces. In contrast, the utilized method generates multi-domain volumetric meshes directly using the input irregular meshes from the generated PDL rims (shown in red in \textbf{C}) and guarantees the interface congruency as well as the mesh conformity as depicted in \textbf{H}.}
    \label{fig:comparison}
\end{figure*}

\begin{figure*}[t]
\centering
    \includegraphics[width=0.95\textwidth]{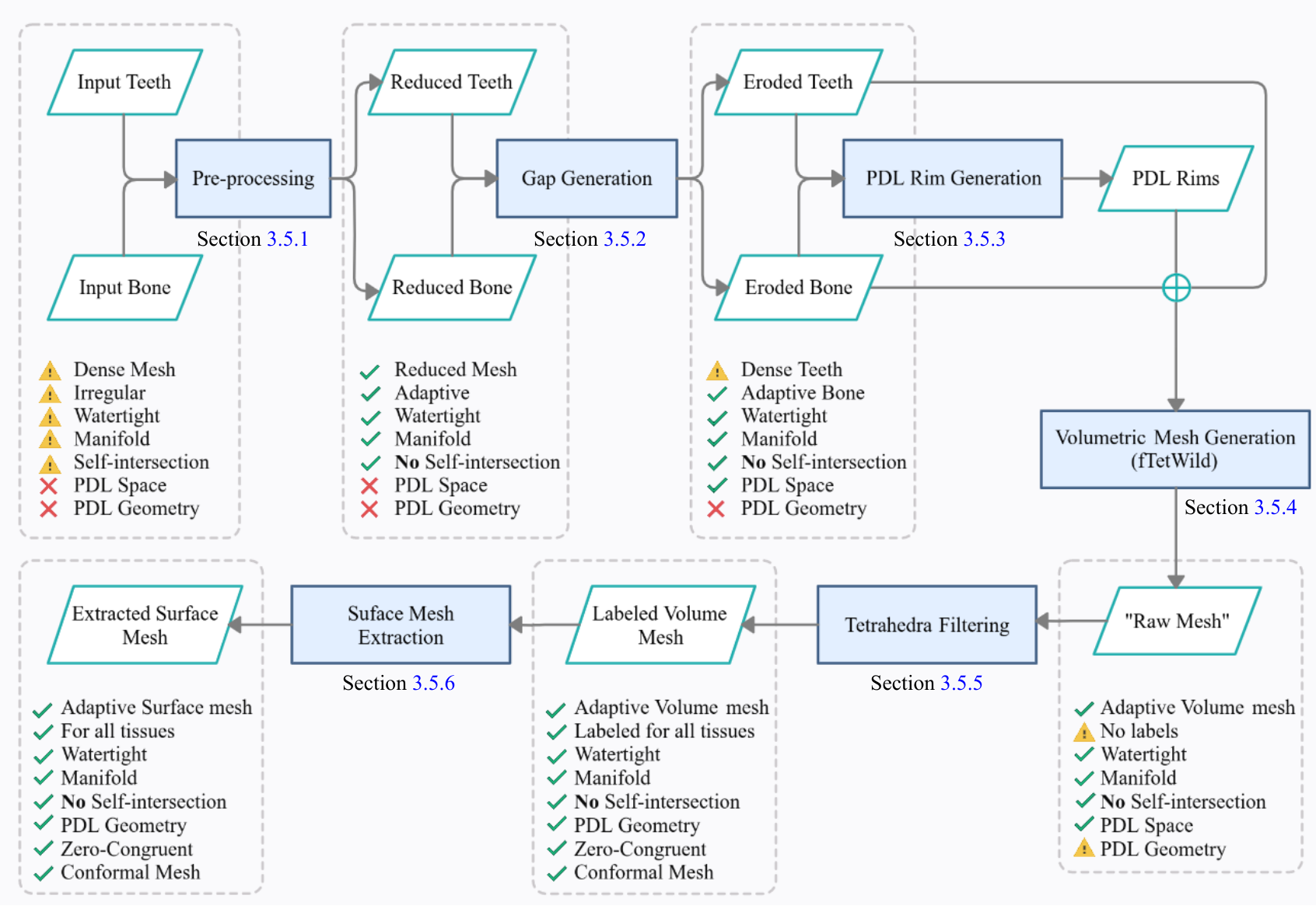}
    \caption{Flowchart of the utilized pipeline and the characteristic of the meshes at each step. The pipeline consists of five main consecutive steps, as further described in Section \ref{section:auto_pipe}.}
    \label{fig:flowchart}
\end{figure*}

To eliminate the manual geometry processing, quality meshing, or mesh decimation steps, we propose to directly use the exported surface meshes as an input to our pipeline. Figure \ref{fig:comparison} shows a comparison between the conventional labor-intensive approach and our method for developing FE models of a human jaw. For the reproducibility purpose, the pipeline is implemented based on the free open-access geometry processing library libigl \cite{libigl} and the meshing algorithm fTetWild \cite{ftetwild}. The pipeline generates conformal volumetric meshes using imperfect meshes exported from the segmentation software with minimal human intervention. Figure \ref{fig:flowchart} shows the flowchart of the utilized pipeline and the characteristics of the meshes at different steps.

\subsubsection{Preprocessing} \label{section:preprocess}
As can be seen in Figure \ref{fig:flowchart}, the input meshes to the pipeline are dense irregular meshes, which are not necessarily guaranteed to be watertight, manifold, and self-intersection-free, referred to as ``triangle soup" in the computer graphics \cite{hu2020fast,hu2018tetrahedral,xu2014signed}. Hence, we apply a preprocessing step to both reduce the mesh sizes and produce meshes that are manifold, watertight, and self-intersection-free. These mesh characteristics assure meaningful values for the utilized signed distance functions in the next geometry processing steps of the used pipeline, i.e., the gap and PDL rim generation steps \cite{xu2014signed}. We use fTetWild \cite{ftetwild} as a robust meshing tool to decimate and ``clean up" the imperfect meshes. To be more specific, the teeth and bone geometries are tetrahedralized separately, and the boundary faces of the resulting tetrahedral meshes are then extracted as the cleaned-up reduced surface meshes to be used as the input meshes to the gap generation step. Further details on the surface mesh extraction process can be found in Section \ref{section:surface_extraction}.

\subsubsection{Gap generation for the PDL tissue} \label{section:gap}
PDL has an average width of 0.2 mm \cite{li2018orthodontic}; its width can approximately be 0.15 mm around the middle third of the root and about 0.21 mm \cite{baron2016relationship,white2014oral} to 0.38 mm \cite{li2018orthodontic} near the root apex and cervical regions.

Reconstructing the PDL layer using CBCT scans obtained in vivo from patients in clinics is a challenging process \cite{hohmann2011influence} as the commonly used voxel dimensions for the CBCT scans, ranging from 0.2 mm to 0.5 mm \cite{cattaneo2021orthodontic}, are not fine enough to capture such a thin structure (roughly 0.2 mm) \cite{dorow2003finite,savignano2019biomechanical}. Although the geometry of the PDL layer can be reconstructed by segmenting it from the micro-CTs acquired in vitro, the x-ray exposure in such scans is extremely high and harmful for the human body, and it is usually obtained from dead specimens. Therefore, we first conduct the teeth-bone segmentation from the CBCT scans and then apply geometry processing techniques to create a gap between the teeth and bone geometries where the PDL can reside with an average thickness of 0.2 mm.

Ideally, to generate the required gaps for PDL geometries, we shrink the bone and teeth, each by 0.1 mm. First, we shrink the bone geometry by 0.1 mm, by moving its mesh points in the reverse direction of per-vertex normal with a magnitude of 0.1 mm, which we call it \textit{explicit shrinking approach}. Before performing any \textit{explicit shrinking process}, the radius curvature is locally computed for bone vertices to identify the sharp and thin structures. This is done by calculating the radius $r$ of the mean curvature $h$ at point $t$ based on $r(t) =$1$/h(t)$, as the reciprocal of the curvature at that point. The radius of the curvature provides useful information about the maximum magnitude that the surface vertices can be moved in the opposite direction of the normals of the surface before a singularity occurs. This means that the movement of the nodes with larger values would cause self-intersection issues and artifacts on the surface mesh. For this reason, evaluating the shrinking limits prior to the bone shrinking process is important. In the case where the shrinking limit is less than 0.1 mm, the pipeline shrinks the bone to its maximum shrinking limit, while more shrinking the teeth to compensate for the total desired gap.

To shrink the teeth, we use an \textit{implicit shrinking approach} based on signed distance functions presented in Equation (\ref{eq:sdf}). The signed distance field contains the boundary of the geometry, i.e., zero iso-surface \cite{osher1988fronts} and the information from different offset surfaces with positive and negative values. The positive offsets/iso-surfaces represent a dilated version of the geometry, while the negative values represent the eroded geometry. Consequently, we use an iso-contour of $\phi_{\Omega} (x) = -0.1$ to shrink the tooth with a magnitude of 0.1 mm. Since the iso-contour is still an implicit representation of the shrunk tooth, we use a contouring method based on a marching cubes algorithm \cite{lorensen1987marching} to convert it to an explicit representation including the vertices and connectivity matrix. As a result, wherever the radius curvature might be less than the desired offset value, the \textit{implicit shrinking approach} can prevent any singularities at the root apexes. It can therefore be used as a more robust shrinking approach, especially for thin and sharp geometries with lower radius curvature values (here, the root apex of incisors). Note that the applied approach to the teeth creates a slightly wider space than the desired gap in the root apexes, which is in line with the clinical studies \cite{mortazavi2016review,white2014oral}.

\subsubsection{Boundary representation of PDL}
\label{section:pdl_rim}
Instead of an explicit representation of PDL which uses closed surface meshes, we use a boundary representation (B-Rep) approach using triangle meshes to describe the PDL domain. This needs to be distinguished from the basis spline (B-Spline) representation. we use B-Rep for describing the PDL domain for three main reasons. First, fTetwild uses the winding number information together with the input surface mesh to generate a volumetric mesh with no prior assumptions such as watertight or closed surfaces. Therefore, a shell surface can be used as a part of the input mesh to represent the domain boundary. Second, since fTetwild uses an implicit algorithm, it enables us to use B-Rep and perform boolean operations on different components to describe the domain boundary \cite{ftetwild}. This is while it is not trivial to correctly define the B-Rep models for complex geometries using the Delaunay-based algorithms such as TetGen \cite{si2015tetgen}. Third, B-Rep models help us to achieve zero-congruent contacting interfaces and avoid numerically small values for $\varepsilon$ produced due to machine/floating-point precision. This, in turn, assures congruency and mesh conformity at the contacting interfaces for modeling the PDL layer as mentioned in Section \ref{traits:pdl}.

The PDL geometry can be described using three main surfaces: the top surface of PDL, tooth-PDL, and PDL-bone interfaces, in which the two last interfaces can be replaced with the teeth and bone geometries. To represent the free surface of the PDLs that are not in contact with the teeth and bone geometries, we generate only the top surface of each PDL, called the PDL rim in this study. B-Rep model using PDL rim and teeth-bone geometries as inputs to fTetWild helps to generate multi-domain volumetric mesh using boolean operations to guarantee geometry congruency and mesh conformity in the contacting interfaces.

We utilize a gap filling method \cite{moshfeghifar2022direct} to generate the rims that connect the bone to shrunk teeth. This method detects an initial surface on the bone within the desired distance like 0.3 mm. Next, the boundary nodes of the detected surface are smoothened using a Laplacian smoothing approach to provide a base surface on the tooth socket with a smooth boundary on the bone. The smoothened boundary points are then snapped to the bone surface to assure the smoothened boundary is on the bone. Afterward, the base function with smoothened boundary is extruded with the desired thickness (0.2 mm), and the extruded surface is snapped to the tooth surface to assure the points are located on the tooth surface. Finally, the PDL rim is constructed by connecting the boundary points of the base and extruded surfaces. We only use the PDL rims without the base/extruded surface to avoid any redundant representations of tooth-PDL and PDL-bone interfaces having potential numerical errors. These errors can cause issues in the volumetric mesh generation process of fTetwild, when the value of the user-defined maximum deviation parameter called envelope is less than or equal to the produced error in the duplicated surfaces.

\begin{figure*}[t]
\centering
    \includegraphics[width=\textwidth]{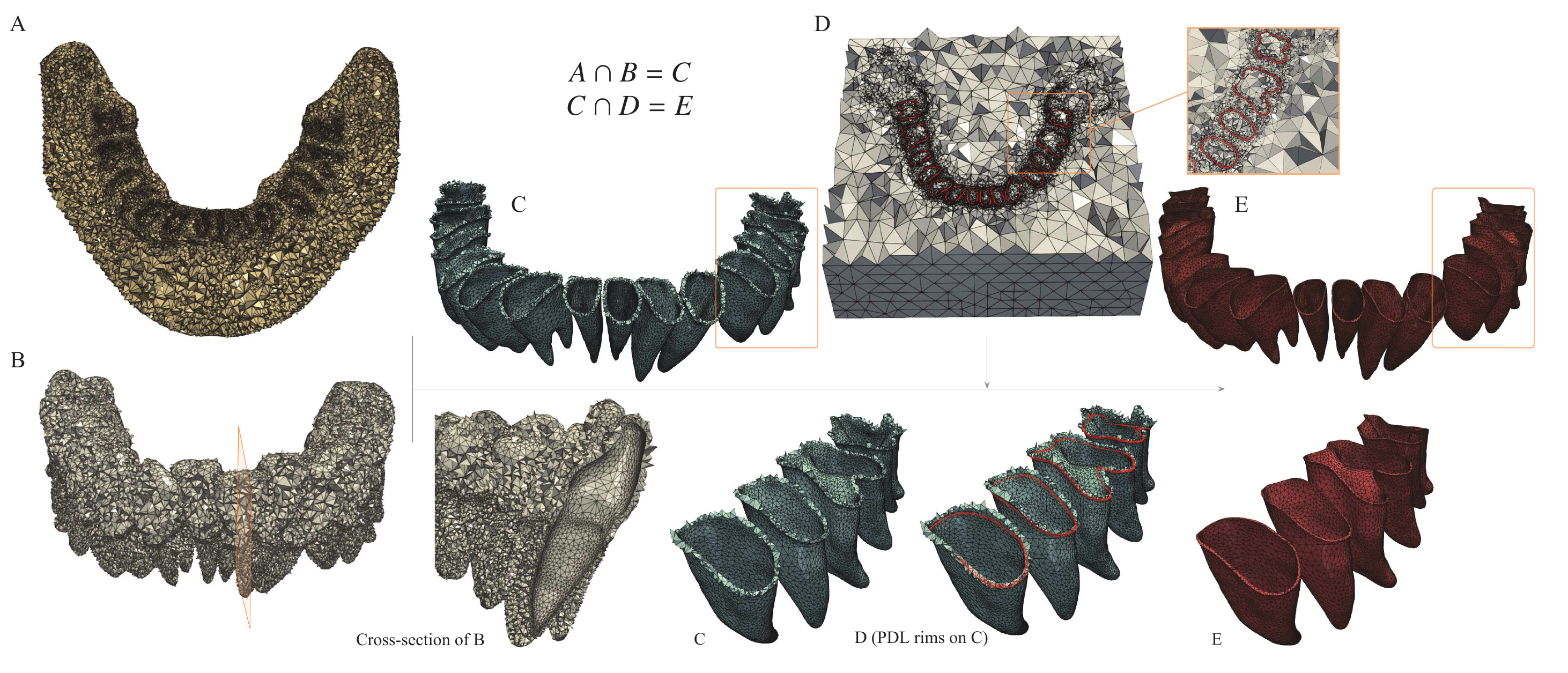}
    \caption{The proposed tetrahedra-filtering method and its close-up view (bottom row) for proper labeling of the PDL from the raw mesh. Thetrahedra around teeth (\textbf{A}) and bone (\textbf{B}) are used to obtain the conformal tooth-PDL and PDL-bone interfaces (\textbf{C}). The obtained mesh includes jagged top surfaces which can be removed by using the positive winding numbers with respect to the PDL rims (\textbf{D}), resulting in smooth and clean top surfaces for the generated PDLs (\textbf{E}).}
    \label{fig:labeling}
\end{figure*}

\subsubsection{Volumetric mesh generation} \label{section:volmesh}

A unified volumetric mesh is generated using the surface meshes of the teeth, bone, and produced PDL rims, It is obtained by applying a union operation on the provided input surface meshes using fTetWild \cite{ftetwild} with optimal \textit{envelope ($\epsilon$)} and \textit{ideal edge length} values of $2 \times 10^{-4}$ and 0.01, respectively. Applying a union operation using fTetWild generates a single volumetric mesh for the bounding box surrounding all input surface meshes called the \textit{raw mesh} \cite{ftetwild}. Note that the default filtering method in fTetWild exports a labeled version of the raw mesh that assigns the tetrahedra outside the closed input surfaces as the background elements.

\subsubsection{Proposed tetrahedra filtering method}
\label{section:tet_filtering}
The default tetrahedra filtering/labeling method in fTetWild can only be used when all input meshes are closed surface meshes, which is not the case for the PDL rims. Hence, a specific filtering approach is used for assigning each tetrahedron a label associated with one of the input domains, i.e., the teeth, PDL, or bone. The remained tetrahedra are labeled as background. We use a distance field-based labeling method for teeth and bone. To do so, the barycenter of each tetrahedron is used as an average point of tetrahedron vertices to decide whether the element is positioned inside or outside a surface considering the distance field of each barycenter with respect to each of the teeth and bone geometries. The negative distance values indicate the points are located inside the surface mesh. 

We propose using a signed distance-based method combined with the winding number information to obtain the PDL geometries. The PDL is expected to reside in the gap between the teeth and bone; hence, as shown in Figure \ref{fig:labeling}, we first select all tetrahedra with positive distance values below 0.3 mm around teeth (A) and bone (B) based on the intersection of these two sets of tetrahedra (C). Next, the intersection of the obtained tetrahedra (C) and the tetrahedra with positive winding numbers with respect to the PDL rims (D) is utilized to produce a smooth boundary (E) for the top-free surface of each PDL. The winding numbers, as also used in fTetWild for implicit meshing, are applied to eliminate all meshing artifacts introduced by the intersection operation (C). As can be seen, applying the proposed filtering method considering the information of the winding numbers to the raw mesh provides smooth PDL top surfaces as well as conformal meshes in the tooth-PDL and PDL-bone interfaces, which include shared nodes in the contacting interfaces and provide perfect adhesions in the contacting interfaces.

\subsubsection{Surface mesh extraction}
\label{section:surface_extraction}
After labeling all tetrahedra in the raw mesh, we select boundary faces of each domain, i.e., the shared faces between the domain and background tetrahedra, and save them as reduced adaptive surface meshes. Note that the extracted surfaces are zero-congruent and conformal meshes with no meshing artifacts.

\subsubsection{FE simulation setup}

The used pipeline automatically sets up the FE problems based on two different scenarios; (1) uncontrolled tipping and (2) biting scenarios. It automatically generates FEM files of the uncontrolled tipping scenario for all patients and FEM files of the biting scenario for those whose scans were acquired in a natural biting position. Note that the utilized code for defining the boundary and loading conditions is scenario-specific, which requires to be adapted for other scenarios. The defined FE problem for each scenario is exported as an XML file including the tetrahedral meshes, types of elements, utilized material models and properties of each domain, as well as boundary and loading conditions.

In these scenarios, we use Tet4 elements for the teeth and bone with negligible deformations. Furthermore, to avoid the locking artifacts \cite{Schneider:2018} that are common when linear tetrahedral elements are used to model large deformations, we apply a quadratic basis to the elements of the PDL layer, to consider Tet10 elements for the PDL domain. We prefer to use an explicit tetrahedral mesh to model the PDL layer \cite{ortun2018approach,hohmann2011influence,lee2018biomechanical} (instead of the cheaper option of using shell elements) as it can more faithfully capture shear, bending, and buckling of the membrane, and at the same time, it simplifies both meshing and simulation, as it does not require special handling for inserting shell elements in a tetrahedral mesh and coupling between the volumetric elastic model and its interaction with the shell. Further, providing a volumetric representation of the PDL layer allows future studies to explore the effects of utilizing visco-elastic-plastic material models for the PDL layer.

\paragraph{Material models and parameters} In this study, the bone and teeth tissues deform negligibly under the applied forces. Therefore, we assume no distinctions between different structures of the bone (cortical and trabecular) and teeth (enamel, dentin, and pulp) \cite{gholamalizadeh2020mandibular,ziegler2005numerical,qian2009deformation}.  Besides, the porous fibrous periodontal ligament tissue is assumed as a homogeneous structure \cite{ortun2018approach}. We use the Neo-Hookean material model with Poisson's ratios of 0.3, 0.3, and 0.45, and Young's modulus values of 2000 MPa, 1500 MPa, and 68.9 MPa to describe the mechanical behavior of the tooth, PDL, and bone domains, respectively \cite{gholamalizadeh2020mandibular,benazzi2013unravelling,benaissa2020stress}.

Note that the used simplex material models can be replaced with any complex constitutive models to properly mimic the anisotropic viscoelastic behavior of the PDL \cite{qian2009deformation,fill2012analytically,ortun2018approach}, or orthotropic characteristic of the bone. This can however increase the computational costs of the simulations. 

\paragraph{Boundary conditions} In the tipping scenario, a Dirichlet boundary condition is defined on all nodes located at the bottom/top surface of the mandible/maxilla to fix displacements of the nodes in all three directions.
In the biting scenario, the same Dirichlet boundary condition is applied on the top nodes of the maxilla to fix them in all three directions. Besides, a similar Dirichlet boundary condition is used to restrict the movement of the mandible only in the anterior-posterior and medial-lateral directions with imposing no restrictions in the third direction.

\paragraph{Loading conditions} In the tipping scenario, we apply a perpendicular force with a magnitude of 1 N at the center of each crown to mimic the uncontrolled tipping motion in the lingual direction. In the biting scenario, a pressure load of 200 N is applied to the bottom surface of the mandible to simulate the biting force. 

\paragraph{Contact definition} To have a perfect adhesion in the tooth-PDL and PDL-bone interfaces, we generate a single volumetric mesh by combining surface meshes of the shrunk teeth, PDL rim, and bone geometries using a union operation in fTetWild \cite{ftetwild}. The nodes at the contacting interfaces are shared between the adjacent domains, thus guaranteeing a complete adhesion as well as conformal meshes at the contacting interfaces. Therefore, there will be no sliding, separation, or penetration at the tooth-PDL and PDL-bone contacting interfaces. Further details on the volumetric mesh generation can be found in Section \ref{section:volmesh}.

Furthermore, in the both scenarios, any potential contacts between different teeth are modeled using the incremental potential contact formulation \cite{Li2020IPC}.

\subsubsection{Technical details}

\paragraph{Implicit shrinking approach} The explicit representation of geometries only provides boundary information of them. This is while the implicit representation of geometries based on the signed distance function provides useful information about the interior and exterior of the shape as well as the outline/boundary of geometry. Therefore, we use boundary information obtained from the explicit representation to define the signed distance function/filed before evaluating it on the desired points in an $R^3$ space and shrinking the teeth. To do so, we sample the distance field in the $R^3$ space using a regular 3D grid in the bounding box of each tooth. A fine grid sampling size of 0.1 mm is used to be smaller than the minimum isotropic voxel size (0.15 mm) as presented in Table \ref{table:scan_info}, to avoid the aliasing effect or spiky surfaces in the sampling process and have smooth geometries in the shrunk teeth. 

Next, the signed distance values are obtained for each point of the grid considering the surface mesh of the tooth. Note that sampling using regular grids requires excessive memory for geometries with large dimensions, which is not the case here as the dimension of each tooth is relatively small. In general, for meshes with fine details that cover a larger space, it is suggested to use adaptive sampling approaches such as the octree-based methods \cite{meagher1982geometric}, to densely sample voxels in specific regions near the boundary or regions with great details.

\paragraph{Alternative PDL modeling} An alternative approach for generating FE models of the tooth-supporting complex can be using Delaunay-based volumetric meshing tools like TetGen. Still, one needs to assure mesh conformity or interface congruency on the surface meshes, and next, to enforce the explicit volumetric meshing algorithm to preserve the surface meshes \cite{vukicevic2021openmandible}. Note that providing quality surface meshes with mesh conformity criteria per se is not trivial. Besides, even though the alternative explicit approach can provide quality surface meshes, it may not provide optimal quality for the tetrahedra due to the additional constraint applied to preserve the surface mesh on the boundary of the domain.

\begin{table}[]
\caption{The thickness of the PDL layers generated using the utilized pipeline with the statistics in line with the literature \cite{li2018orthodontic}. Note that there is a significant difference between our model results and those of the OpenMandible.}
\vspace{1.5mm}
\resizebox{\columnwidth}{!}{%
\begin{threeparttable}
\begin{tabular}{lcccccc}
\hline 
\multicolumn{1}{c}{\multirow{2}{*}{\begin{tabular}[c]{@{}c@{}}Patient's~~\\ ID\end{tabular}}} & \multicolumn{3}{c}{Mandibular jaw} & \multicolumn{3}{c}{Maxillary jaw} \\ \cline{2-7}
& Minimum & Maximum & Average & Minimum & Maximum & Average \\ \hline
\rowcolor{Gray}
Patient 1 & 0.12316 & 0.23552 & 0.19530 & - & - & -  \\
Patient 2 & 0.14227 & 0.25042 & 0.19536 & - & - & -  \\
\rowcolor{Gray}
Patient 3 & 0.13779 & 0.33969 & 0.19379 & 0.08258 & 0.38502 & 0.20143 \\
Patient 4 & 0.13779 & 0.21958 & 0.11962 & 0.14331 & 0.23754 & 0.19634 \\
\rowcolor{Gray}
Patient 5 & 0.14537 & 0.24688 & 0.19770 & 0.11837 & 0.23631 & 0.19676 \\
Patient 6 & 0.16350 & 0.24151 & 0.19737 & 0.14928 & 0.25741 & 0.19654 \\
\rowcolor{Gray}
Patient 7 & 0.15258 & 0.25176 & 0.19527 & - & - & -  \\
Patient 8 & 0.13036 & 0.25292 & 0.20056 & 0.15210 & 0.32516 & 0.19745 \\
\rowcolor{Gray}
Patient 9 & 0.13324 & 0.27437 & 0.19629 & - & - & -  \\
Patient 10 & 0.11951 & 0.23903 & 0.20100 & - & - & -  \\
\rowcolor{Gray}
Patient 11 & 0.14271 & 0.33307 & 0.19887 & 0.12834 & 0.25599 & 0.19465 \\
Patient 12 & 0.11694 & 0.34803 & 0.20164 & 0.12070 & 0.31058 & 0.19219 \\
\rowcolor{Gray}
Patient 13 & 0.12288 & 0.28818 & 0.20134 & 0.09648 & 0.26672 & 0.18296 \\
Patient 14 & 0.12056 & 0.29998 & 0.19728 & 0.10014 & 0.29037 & 0.19222 \\
\rowcolor{Gray}
Patient 15 & 0.13099 & 0.35793 & 0.20078 & 0.13277 & 0.34097 & 0.19148 \\
Patient 16 & 0.12508 & 0.25914 & 0.19961 & 0.14156 & 0.30228 & 0.19535 \\
\rowcolor{Gray}
Patient 17 & 0.13754 & 0.26163 & 0.19664 & 0.12427 & 0.28902 & 0.19610 \\ \hline
\begin{tabular}[c]{@{}l@{}}Mean\\ $\pm$ STD\end{tabular} &
\begin{tabular}[c]{@{}c@{}}0.13425\\ $\pm$ 0.0126\end{tabular} & \begin{tabular}[c]{@{}c@{}}0.27645 \\ $\pm$ 0.04360\end{tabular} & \begin{tabular}[c]{@{}c@{}}0.19344\\ $\pm$ 0.01918\end{tabular} & \begin{tabular}[c]{@{}c@{}}0.12416\\ $\pm$ 0.02193\end{tabular} & \begin{tabular}[c]{@{}c@{}}0.29145\\ $\pm$ 0.044527\end{tabular} & \begin{tabular}[c]{@{}c@{}}0.19446\\ $\pm$ 0.0045\end{tabular} \\ \hline
OpenMandible & 0.20576 & 1.93415 & 0.60772 & - & - & -  \\ \hline
\end{tabular}%
\vspace{0.5mm}
\label{table:pdl_info}
\end{threeparttable}
}
\end{table}

\begin{figure*}[t]
\centering
    \includegraphics[width=\textwidth]{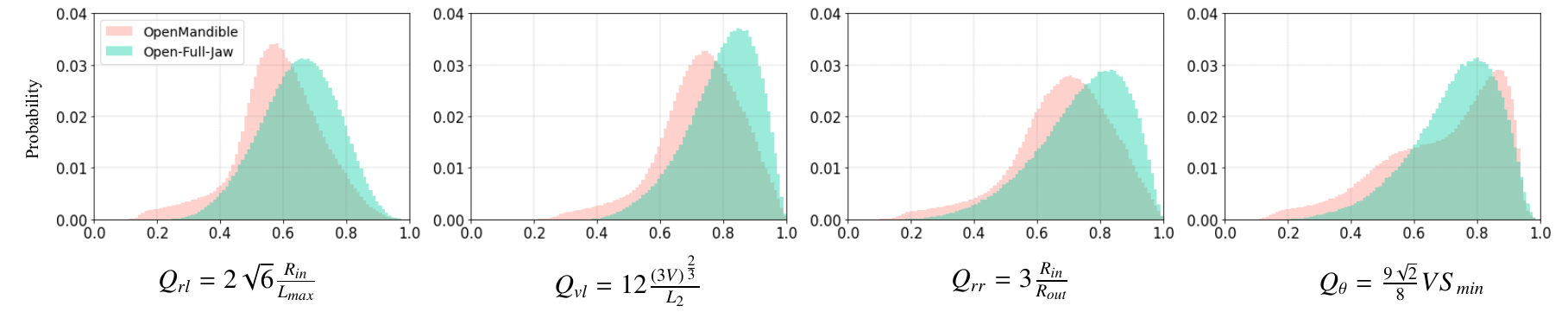}
    \caption{Comparison of the quality histograms of four different quality measurements used for the evaluation of the quality of the tetrahedra generated by OpenMandible and Open-Full-Jaw. \textbf{Left to right:} The radius-edge ratio $Q_{rl}$; The volume-edge ratio $Q_{vl}$; The radius ratio $Q_{rr}$; The angle measurement $Q_{\theta}$. Note that $L_{max}$, $R_{in}$, and $R_{out}$ denote the maximum edge length, the radius of the insphere, and radius of the circumsphere of a tetrahedron, respectively, and $V$ and $S_{min}$ represent the volume and minimum face area of the tetrahedron.}
    \label{fig:hists}
\end{figure*} 

\paragraph{Advantages of implicit over explicit meshing algorithms} The explicit volumetric mesh generation approaches such as the Delaunay-based algorithms with incremental local mesh operations generate tetrahedral meshes covering the domain interior and are highly faithful to the input mesh. This means that the algorithms cannot coarsen the triangles on the input surface meshes. Therefore, they cannot produce coarse quality tetrahedra on the boundary faces of the input mesh, indicating that the element size at the domain boundary depends on the dense and irregular/rugged surfaces exported from the segmentation step. Consequently, one needs to prepare an adaptive surface mesh with high-quality triangles before applying the Delaunay-based algorithms along with a sizing field function to create adaptive volumetric mesh both in the domain interior and its boundary. Furthermore, since the explicit meshing algorithms fail in generating volumetric meshes in the presence of any self-intersections, one needs to clean up the meshes before applying the Delaunay-based meshing algorithms to create quality and adaptive mesh.

In contrast, implicit volumetric meshing approaches like fTetWild impose no assumptions on the input mesh and can handle imperfect input meshes with self-intersection artifacts. They can also produce an adaptive mesh that provides coarse tetrahedra at the domain boundary, slightly deviating from the input surface mesh, based on a user-defined input parameter. In addition, fTetWild uses winding numbers along with surface meshes to generate tetrahedral meshes based on an implicit meshing approach. This enables us to use the constructive solid geometry (CSG) model for describing a domain with complex boundaries by combining several simpler domains using boolean operations \cite{foley1996computer}. This is while the explicit meshing tools cannot support CSG models \cite{si2015tetgen}.

\paragraph{Volumetric meshing} The fTetWild algorithm uses the user-defined epsilon and ideal edge length parameters to control the output mesh's accuracy and size. The epsilon value indicates how much fTetWild can deviate from the input surface mesh for generating a quality volumetric mesh. Hence, using a smaller epsilon value preserves the input geometry's details at the cost of higher computational time. In contrast, larger values can provide less accurate meshes in less computational time. Moreover, smaller values of ideal edge length provide denser meshes, while larger values can produce coarser elements. Hence, these two parameters can control the output mesh size and its geometrical accuracy. In this work, the utilized values for epsilon and ideal edge length are obtained based on a grid-search method in the intervals $[10^{-6}, 10^{-3}]$ and $[0.005, 0.02]$, respectively. Note that the fTetWild multiplies these parameters by the length of the diagonal of the input meshes' bounding box. Hence, the parameters are sensitive to the size of the bounding box that includes all input meshes. In other words, they need to be adjusted according to the model's size; if, for instance, the model is a cut-out model with a much smaller bounding box dimension than the full jaw model.

\section{Results and discussion}
In this section, the generated models are assessed from different aspects, and the results are compared with the state-of-the-art. In Section \ref{section:pdl_thickness}, we measure the thickness of the generated PDL layers. In Section \ref{section:mesh_properties}, different mesh properties such as mesh sizes and mesh qualities are evaluated for the developed models; the results are compared to the OpenMandible model. In Section \ref{section:FE_result}, FE simulation results of different patients are presented under identical tipping and biting scenarios.

\subsection{Generated PDL properties} \label{section:pdl_thickness}
We assume an average thickness of 0.2 mm for the PDL layers according to the literature \cite{seo2021comparative,ortun2020silico} and assess the thicknesses of the generated PDLs across different patients models by computing the distances between the points located on the outer and inner surfaces of the PDL geometries. The thickness details of the generated PDLs can be seen in Table \ref{table:pdl_info}. As can be noticed from the table, the computed minimum, maximum, and average PDL thickness values are in line with those reported in the literature, i.e., 0.15, 0.3, and 0.2mm, respectively \cite{li2018orthodontic}.

Likewise, we investigate the thicknesses for the PDLs from the OpenMandible dataset. As it can be seen in the last row of Table \ref{table:pdl_info}, the obtained thicknesses for OpenMandible are about two-three times the values reported in the literature. As the PDL thickness plays an essential role in exerting the applied load from teeth surfaces to the adjacent bone and teeth movements, it can be deduced that under identical scenarios and loading systems the OpenMandible model cannot result in stress/strain values comparable with the ones obtained in this study.

\subsection{Mesh properties}
\label{section:mesh_properties}
We evaluate the total number of elements and mesh qualities of the developed models and compare the mesh qualities of our model with those of the OpenMandible dataset. The maximum mesh deviation between the input and output meshes of ftetwild is then calculated to examine whether the meshing process alters the geometries negligibly.

\subsubsection{Mesh sizes}
We develop all computational meshes of our study on a machine with a 2.60 GHz processor and 16 GB of RAM, which takes around $40.78 \pm 12.19$ minutes per jaw. The details of the original dense meshes, the generated volumetric meshes, and their extracted surface meshes are summarized in Table \ref{table:mesh_sizes}. The pipeline produces conformal adaptive meshes from irregular dense meshes. The utilized adaptive meshing approach helps to reduce the total number of the elements, by producing coarser elements in regions far from the teeth while generating finer elements on root apexes, PDL layer, alveolar crest, or regions with fine details like thin walls of the maxilla. 

\begin{table*}[]
\centering
\caption{The total number of elements ($\times 10^3$) of the surface meshes (triangles) and volumetric meshes (tetrahedra). The number of elements is significantly reduced in the adaptive output surface meshes compared to that of the irregular dense input surface meshes.}
\vspace{0.8mm}
\label{table:mesh_sizes}
\resizebox{\textwidth}{!}{%
\begin{tabular}{cccccccccccccccccccccccc}
\hline
\multirow{3}{*}{\begin{tabular}[c]{@{}c@{}}Patient's\\ ID\end{tabular}} & \multicolumn{11}{c}{Mandibular jaw's mesh sizes} &  & \multicolumn{11}{c}{Maxillary jaw's mesh sizes} \\ \cline{2-12} \cline{14-24} 
 & \multicolumn{2}{c}{Input irregular mesh} &  & \multicolumn{3}{c}{Output surface mesh} &  & \multicolumn{4}{c}{Output volumetric mesh} &  & \multicolumn{2}{c}{Input mesh size} &  & \multicolumn{3}{c}{Output surface mesh} &  & \multicolumn{4}{c}{Output volumetric mesh } \\ \cline{2-3} \cline{5-7} \cline{9-12} \cline{14-15} \cline{17-19} \cline{21-24} 
 & Teeth & Mandible &  & Teeth & PDLs & Mandible &  & Teeth & PDLs & Mandible & Total &  & Teeth & Maxilla &  & Teeth & PDLs & Maxilla &  & Teeth & PDLs & Maxilla & Total \\ \hline
\rowcolor{Gray}
Patient 1 & 198 & 478 &  & 59 & 60 & 75 &  & 178 & 87 & 252 & 517 &  & - & - &  & - & - & - &  & - & - & - & - \\
Patient 2 & 531 & 1574 &  & 50 & 45 & 70 &  & 141 & 65 & 224 & 430 &  & - & - &  & - & - & - &  & - & - & - & - \\
\rowcolor{Gray}
Patient 3 & 900 & 1576 &  & 82 & 77 & 96 &  & 243 & 111 & 315 & 669 &  & 948 & 1311 &  & 100 & 96 & 132 &  & 293 & 140 & 421 & 853 \\
Patient 4 & 770 & 1592 &  & 94 & 108 & 106 &  & 294 & 155 & 338 & 788 &  & 824 & 1894 &  & 78 & 77 & 153 &  & 231 & 112 & 440 & 783 \\
\rowcolor{Gray}
Patient 5 & 768 & 1553 &  & 77 & 88 & 98 &  & 227 & 127 & 326 & 681 &  & 851 & 2039 &  & 84 & 93 & 159 &  & 249 & 134 & 496 & 880 \\
Patient 6 & 731 & 1406 &  & 78 & 82 & 91 &  & 233 & 119 & 312 & 664 &  & 904 & 2181 &  & 89 & 92 & 173 &  & 250 & 134 & 502 & 886 \\
\rowcolor{Gray}
Patient 7 & 526 & 1556 &  & 51 & 49 & 73 &  & 142 & 71 & 236 & 448 &  & - & - &  & - & - & - &  & - & - & - & - \\
Patient 8 & 625 & 2000 &  & 50 & 54 & 83 &  & 150 & 78 & 274 & 502 &  & 772 & 1921 &  & 75 & 70 & 157 &  & 221 & 101 & 469 & 791 \\
\rowcolor{Gray}
Patient 9 & 869 & 2110 &  & 74 & 76 & 99 &  & 230 & 110 & 341 & 681 &  & - & - &  & - & - & - &  & - & - & - & - \\
Patient 10 & 791 & 1867 &  & 69 & 71 & 91 &  & 202 & 103 & 301 & 606 &  & - & - &  & - & - & - &  & - & - & - & - \\
\rowcolor{Gray}
Patient 11 & 708 & 2201 &  & 53 & 51 & 79 &  & 160 & 74 & 264 & 497 &  & 891 & 1531 &  & 91 & 85 & 135 &  & 260 & 124 & 416 & 799 \\
Patient 12 & 233 & 629 &  & 57 & 60 & 82 &  & 163 & 87 & 261 & 511 &  & 232 & 636 &  & 67 & 61 & 133 &  & 203 & 87 & 399 & 689 \\
\rowcolor{Gray}
Patient 13 & 188 & 469 &  & 54 & 50 & 78 &  & 163 & 72 & 252 & 487 &  & 190 & 1016 &  & 52 & 50 & 191 &  & 156 & 72 & 548 & 776 \\
Patient 14 & 183 & 486 &  & 50 & 49 & 72 &  & 150 & 71 & 238 & 459 &  & 189 & 852 &  & 50 & 50 & 139 &  & 141 & 72 & 394 & 607 \\
\rowcolor{Gray}
Patient 15 & 1016 & 1969 &  & 85 & 101 & 108 &  & 258 & 149 & 371 & 777 &  & 1032 & 2221 &  & 90 & 108 & 175 &  & 261 & 159 & 519 & 939 \\
Patient 16 & 877 & 2958 &  & 61 & 69 & 97 &  & 194 & 99 & 311 & 605 &  & 851 & 2958 &  & 67 & 77 & 166 &  & 209 & 110 & 467 & 786 \\
\rowcolor{Gray}
Patient 17 & 440 & 1045 &  & 69 & 80 & 94 &  & 210 & 116 & 305 & 630 &  & 447 & 577 &  & 91 & 93 & 126 &  & 273 & 135 & 378 & 787 \\ \hline
\end{tabular}%
}
\end{table*}

\subsubsection{Mesh quality analysis}
The mesh quality required for an FE analysis can vary depending on the application and utilized numerical methods \cite{shewchuk2002good}. In general, a regular tetrahedron has the highest mesh quality in computational models, and the main guideline is to avoid using low-quality/badly-shaped tetrahedra, as they can affect the accuracy of the numerical methods.

The OpenMandible uses TetGen to obtain volumetric meshes from the manually generated conformal surface meshes. It sets the upper limit of the radius-edge ratio of to-be-generated tetrahedra to 1.5. This mesh quality constraint controls the ratio between the radius of the circumscribed sphere and the shortest edge of each tetrahedron, to prevent the production of low-quality/badly-shaped tetrahedra. In addition to this quality constraint, OpenMandible enforces TetGen to preserve the provided input surface meshes to have conformal volumetric meshes. Preserving the surface mesh is the only approach to produce conformal volumetric meshes when using explicit volumetric meshing approaches. This raises the question of whether TetGen can achieve the specified quality constraint value while enforcing another restriction to preserve the input surface mesh.

We quantitatively assess the quality of the generated volumetric meshes of this study and those of the OpenMandible. To be more specific, four different quality measurements presented in \cite{shewchuk2002good}, i.e., the radius-edge ratio $Q_{rl}$ \cite{baker1989element}, the volume-edge ratio $Q_{vl}$ \cite{liu1994relationship,misztal2013multiphase}, the radius ratio $Q_{rr}$ \cite{freitag1997tetrahedral,caendish1985apporach}, and the angle measurement $Q_{\theta}$ \cite{freitag1997tetrahedral}, are used to evaluate the quality of the volumetric meshes. For a fair comparison, we compare the results from one of our patients to those of the OpenMandible study, as shown in Figure \ref{fig:hists}.

In a regular tetrahedron, the values of each utilized quality measurement correspond to one, indicating that an ideal high-quality mesh is expected to have a histogram peak at one. Therefore, the resulting distributions (normal or skewed normal) of all quality histograms show that our generated volumetric mesh has higher quality elements compared to OpenMandible. The proposed model also has narrower distributions with small discrepancies around their means. This holds even in the last case ($Q_{\theta}$), where one of the peaks of the distribution (mode) for OpenMandible is closer to one. Moreover, the OpenMandible model sees two or more peaks in its distributions that can be modeled by mixed normal distributions.

\begin{table}[!ht]
\centering
\caption{Hausdorff distances (HD) used as an error measurement to assess the deviation of the surface in the output mesh from the input mesh. The HD values indicate that the output surface meshes generated using our pipeline are close to the input surface meshes.}
\vspace{2mm}
\label{table:hd_dist}
\resizebox{\columnwidth}{!}{%
\begin{tabular}{lcccc}
\hline
\multicolumn{1}{c}{\multirow{2}{*}{\begin{tabular}[c]{@{}c@{}}Patient's\\ ID\end{tabular}}} & \multicolumn{2}{c}{\begin{tabular}[c]{@{}c@{}}Mandibular Jaw HD\\  (mm)\end{tabular}} & \multicolumn{2}{c}{\begin{tabular}[c]{@{}c@{}}Maxillary Jaw HD\\  (mm)\end{tabular}} \\ \cline{2-5} 
\multicolumn{1}{c}{} & Mandible & Mandibular Teeth & Maxilla & Maxillary Teeth \\ \hline
\rowcolor{Gray}
Patient 1 & 0.173 & 0.038 & - & - \\
Patient 2 & 0.147 & 0.039 & - & - \\
\rowcolor{Gray}
Patient 3 & 0.125 & 0.044 & 0.143 & 0.019 \\
Patient 4 & 0.086 & 0.043 & 0.137 & 0.029 \\
\rowcolor{Gray}
Patient 5 & 0.139 & 0.025 & 0.156 & 0.018 \\
Patient 6 & 0.241 & 0.023 & 0.232 & 0.042 \\
\rowcolor{Gray}
Patient 7 & 0.154 & 0.042 & - & - \\
Patient 8 & 0.454 & 0.029 & 0.135 & 0.039 \\
\rowcolor{Gray}
Patient 9 & 0.174 & 0.043 & - & - \\
Patient 10 & 0.453 & 0.021 & - & - \\
\rowcolor{Gray}
Patient 11 & 0.152 & 0.027 & 0.298 & 0.023 \\
Patient 12 & 0.162 & 0.024 & 0.212 & 0.044 \\
\rowcolor{Gray}
Patient 13 & 0.158 & 0.022 & 0.320 & 0.024 \\
Patient 14 & 0.211 & 0.026 & 0.349 & 0.042 \\
\rowcolor{Gray}
Patient 15 & 0.205 & 0.033 & 0.231 & 0.020 \\
Patient 16 & 0.189 & 0.046 & 0.404 & 0.032 \\
\rowcolor{Gray}
Patient 17 & 0.180 & 0.028 & 0.257 & 0.016 \\ \hline
\multicolumn{1}{c}{\begin{tabular}[c]{@{}c@{}}Mean\\  $\pm$ STD\end{tabular}} & \begin{tabular}[c]{@{}c@{}}0.200\\  $\pm$ 0.102\end{tabular} & \begin{tabular}[c]{@{}c@{}}0.032\\  $\pm$ 0.009\end{tabular} & \begin{tabular}[c]{@{}c@{}}0.240\\  $\pm$ 0.089\end{tabular} & \begin{tabular}[c]{@{}c@{}}0.029\\  $\pm$ 0.011\end{tabular} \\ \hline
\end{tabular}%
}
\end{table}

\begin{figure*}[t]
\centering
    \includegraphics[width=\textwidth]{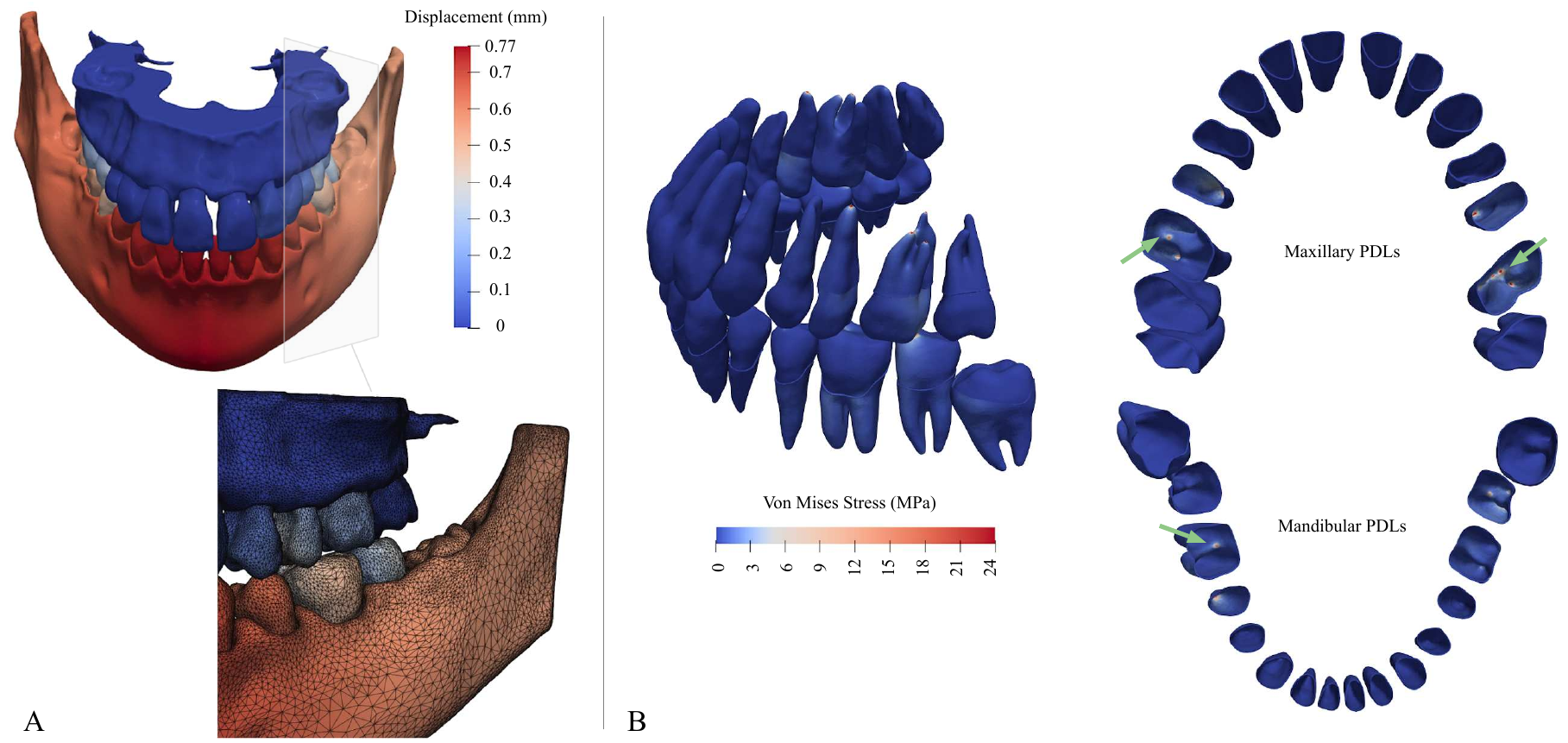}
    \caption{The obtained displacement (\textbf{A}) and stress (\textbf{B}) fields for Patient 17 under the biting scenario. Note that due to the deformation of the PDL layer caused by the exerted biting force from the contacting mandibular teeth, the displacement field is seen on the posterior maxillary teeth (\textbf{A}). Likewise, stress concentrations can be seen on the associated PDL layers, especially at the apexes and furcation regions, indicated by arrows in \textbf{B}.}
    \label{fig:result_biting}
\end{figure*}

\begin{figure*}[!ht]
\centering
    \includegraphics[width=\textwidth]{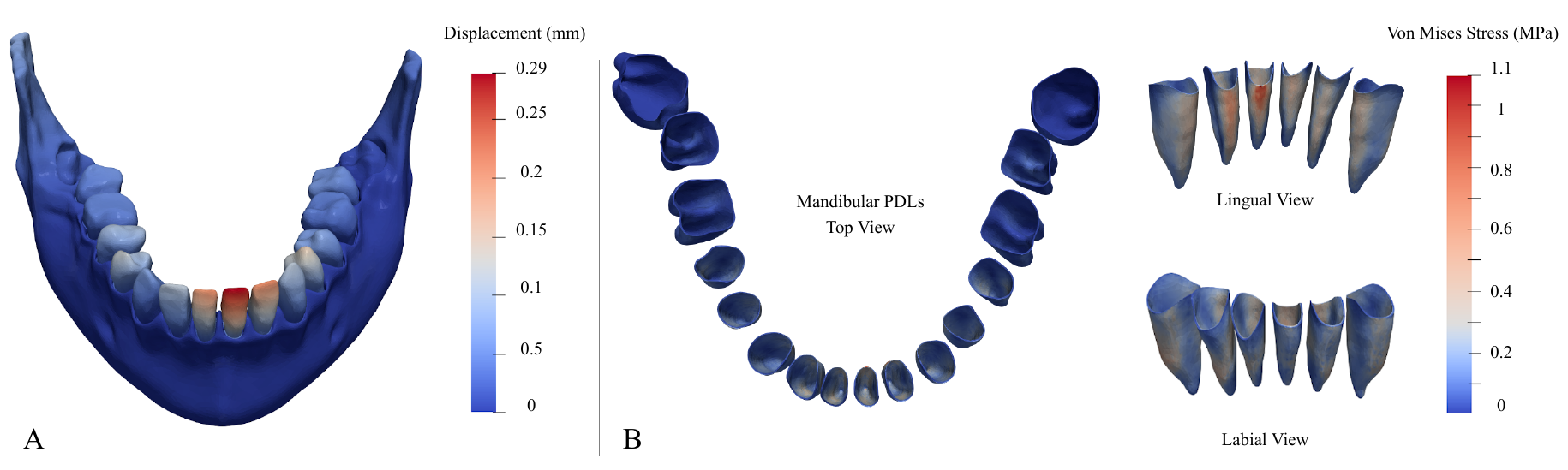}
    \caption{The obtained displacement (\textbf{A}) and stress (\textbf{B}) fields for the mandibular jaw of Patient 17 under the tipping scenario. Note that stress concentration is observed on the lingual side of anterior teeth due to the labiolingual tipping load.}
    \label{fig:result_tipping}
\end{figure*}

\subsubsection{Mesh deviation analysis}
If one does not enforce the surface preservation criterion, the Delaunay-based volumetric meshing algorithms like TetGen include all points of the input surface mesh and a number of additional points. Hence, the generated volumetric meshes have boundary elements as fine as the input surface mesh. In contrast, fTetwild provides volumetric meshes as coarse as possible on the surface while preserving the input geometry. This algorithm slightly deviates from the input surface according to the user-defined maximal deviation (envelope) value. Therefore, to generate an accurate volumetric mesh, we must not deviate too much from the input surface mesh.

Accordingly, we quantitatively evaluate the deviation of final extracted surface meshes from the input surface meshes. We use Hausdorff distance as an error measurement between the input and extracted surface meshes from the generated volumetric meshes. The obtained distance values are presented in Table \ref{table:hd_dist}. As it can be seen, the maximum deviation of the meshes both in mandibular and maxillary jaws and their associated teeth is negligible with respect to the dimension of the entire jaw models.

\subsection{FEM verification} \label{section:FE_result}

As the last part of our analyses, we assess the displacement and stress fields in the tipping and biting scenarios to ensure that the stress patterns are smooth and have no unrealistic stress concentrations. To do so, we run the simulations using the PolyFEM \cite{polyfem} FE solver. PolyFEM is an FE simulation toolkit that supports elastodynamic deformations with linear and non-linear material models. It provides an adaptive p-refinement that allows increasing the order of basis functions for specific domains while utilizing linear basis functions for the other domains. Hence, we use Tet10 elements for the PDL layer to increase the simulations' accuracy and avoid element locking issues \cite{Schneider:2018:DSA}. Besides, PolyFEM uses the incremental potential contact formulation \cite{Li2020IPC} for contact response and friction, which ensures valid, penetration-free meshes during the entire simulation. The contacts are automatically detected by proximity; hence, there is no need to specify contact surfaces, which significantly simplifies the scene setup.

The simulation results are visualized in ParaView \cite{ahrens2005paraview}. Figure \ref{fig:result_biting} shows the result of the biting scenario of a selected patient, including the resulting displacement and stress fields. Due to the deformation of the PDL layer under the applied biting load, the displacement fields can be seen both on the mandibular jaw and the posterior maxillary teeth (A). Besides, stress concentrations can be observed on the associated PDL layers, especially at root apexes and bifurcation regions indicated by arrows in (B).

Figure \ref{fig:result_tipping} illustrates the result of the auto-generated uncontrolled tipping scenario for the mandibular jaw of the same patient. As seen in \cite{gholamalizadeh2020mandibular}, the anterior teeth have higher displacement fields than the posterior teeth under an identical load (A). Hence, higher stress concentrations can be seen on the lingual side of the PDLs associated with the anterior teeth (B). Additionally, simulation results of different patients in the tipping scenario are shown in Figure \ref{fig:result_tipping_all}. As can be noticed, the stress values considerably change from one patient to another, indicating the importance of utilizing population models for multi-patient analysis.

It should be noted that although we have tested our models under two scenarios, the developed volumetric meshes can be used in various scenarios and different FE frameworks. Furthermore, the FE models can benefit from using more complex material models, boundary, and loading conditions. For example, the provided meshes can be integrated with outputs of other studies \cite{vukicevic2021openmandible,ortun2020silico} to consider masticatory muscles for more realistic biting scenarios.

\section{General discussion}

The utilized and conventional meshing approaches generate the volumetric meshes using reconstructed geometries based on accurately segmented scans. However, obtaining such an accurate segmentation is inherently time-consuming and labor-intensive and, in some cases, could be highly challenging due to the complexity of the problem or lack of high-resolution scans \cite{zheng2020anatomicallyCBCT}. The template-based deformation techniques \cite{liang2017machine,pak2021distortion} can be used to automatically reconstruct the 3D geometries by creating a template mesh and deforming it according to the new data samples. Still, deforming a template mesh using registration approaches or deep learning methods requires accurate spatial registration and high-quality volumetric meshes with no distorted elements.

Obtaining accurate spatial registration and high-quality volumetric meshes for the human jaw can be challenging, as there are large variations among geometries of different patients (Figure \ref{fig:gallery}), such as geometrical differences in the bones and teeth, missing teeth, and topological changes in the number of roots, e.g., the mandibular and maxillary molars with two to four roots. Therefore, to include different types of variations in the data for a plausible deformation, one needs to have different templates covering missing teeth or various numbers of roots. This in turn is a time-consuming process and can increase the complexity of the model. On the other hand, large deformations in small volumes of teeth and roots can result in distorted elements, preventing us from generating high-quality meshes, especially in the PDL layers with thin structures that need to be modeled with fine volumetric elements.

The current study introduces the largest-ever dataset of patient-specific human jaws reconstructed from CBCT scans. We believe this unique clinically validated dataset would pave the way for future population studies in the field. More specifically, data augmentation techniques using machine learning \cite{liang2017shape,pascoletti2021stochastic} can be applied to the Open-Full-Jaw dataset to expand its size and variability by generating plausible synthetic data. In addition, this would enable us to use deep learning methods, which require a large amount of data for training. Still, one needs to use a dataset with enough variations for sampling and assess the generated samples' validity.

\afterpage{
\begin{landscape}
\begin{figure}[!ht]
\centering
    \includegraphics[width=\hsize]{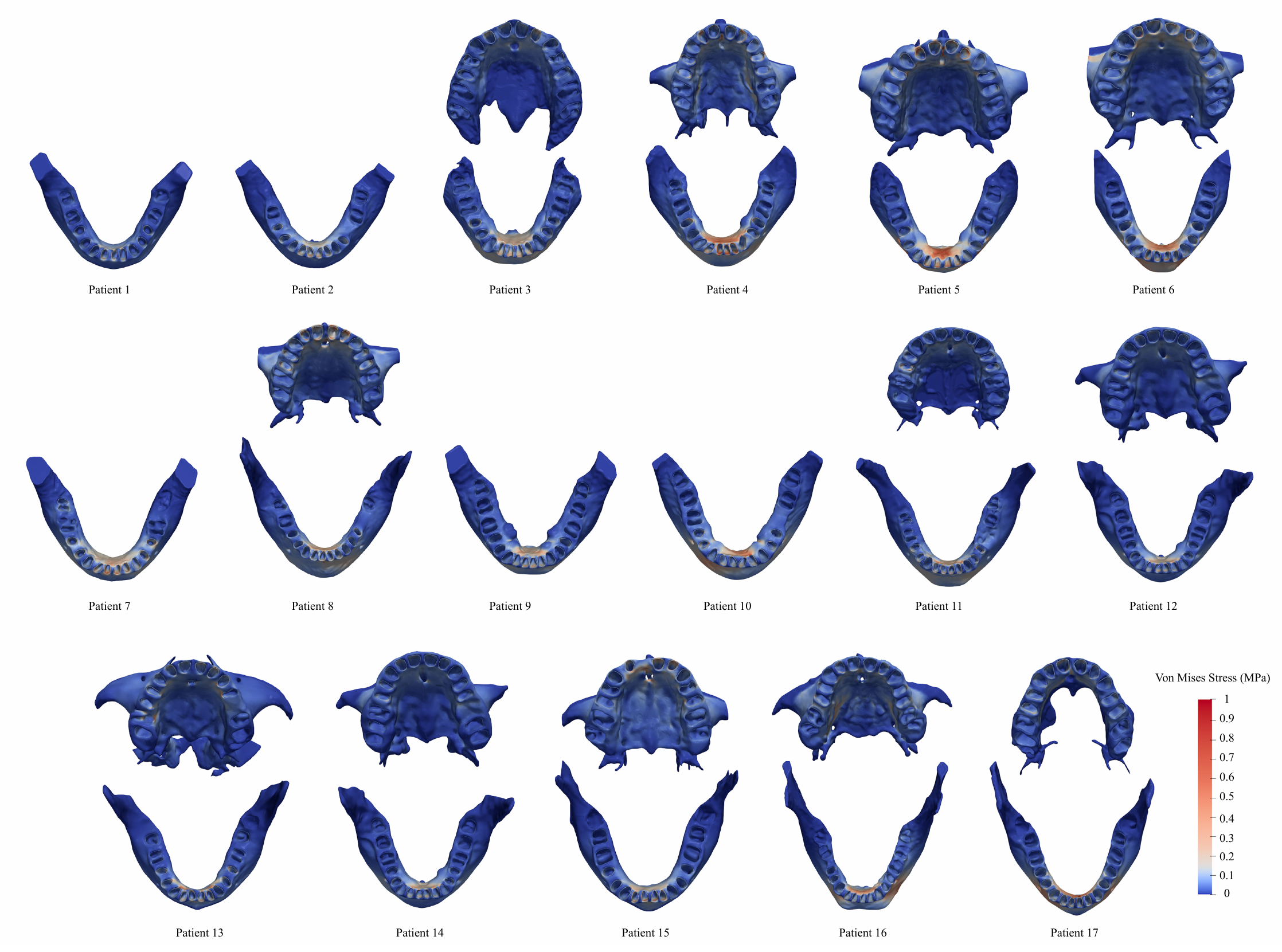}
    \caption{A comparison of stress distributions of all utilized jaws in the tipping scenario with identical load magnitudes. The stress fields are clipped within the range [0, 1] MPa to represent the changes better using the same color map. As can be noticed, the stress values and concentrations considerably vary across different patients, which indicates the importance of utilizing population models for multi-patient analysis and generalizability in computational modeling studies.}
    \label{fig:result_tipping_all}
\end{figure}
\end{landscape}
} 

\section{Conclusion}

In this work, we presented a large open-access dataset, called Open-Full-Jaw (\url{https://github.com/diku-dk/Open-Full-Jaw}), with patient-specific models of 17 human mandibles and maxillae. The dataset contains clinically validated segmented geometries shared as dense surface meshes and adaptive quality volumetric with conformal meshes in the contacting interfaces. It also includes the principal axes for each patient's teeth and the generated FEM files of the uncontrolled tipping and biting scenarios for all patients. Finally, we share the nearly-automated pipeline used for geometry processing, re-meshing, and generating volumetric meshes.

In addition, we evaluated the generated models and quantified them in terms of the mesh quality and accuracy of the models, and compared the results with the state-of-the-art. The obtained results indicate that the developed computational models are precise, considering the low error/distance from input surface meshes. Moreover, the quality of the volumetric elements evaluated based on different quality measurements imply that the generated volumetric meshes consist of quality elements suitable for the FEM of the human jaw. Hence, we believe the Open-Full-Jaw dataset can be used in various FE scenarios and a wide range of intra- and inter-patient analyses.

The shared repository includes all detailed information for reproducing the models of this study. In addition, the utilized pipeline allows other researchers in the field to generate quality volumetric meshes and FE model files directly using dense and irregular meshes with minimal human intervention. This will help other researchers easily extend their datasets without spending much time and effort on manually cleaning up the meshes and non-trivially producing conformal meshes. Furthermore, similar concepts as those used in this study to generate population models of the tooth-supporting complex can be adapted to other areas, such as pelvic girdles and hip joints \cite{libhip}.

\section*{Funding}

This project has received funding from the European Union’s Horizon 2020 research and innovation program under the Marie Sklodowska-Curie grant agreement No. 764644. This paper only contains the authors’ views, and the Research Executive Agency and the Commission are not responsible for any use that may be made of the information it contains. 3Shape A/S provided financial support in the form of salaries for authors TG and PS. The funders had no role in study design, data collection/analysis, publication decision, or manuscript preparation.

This work was also partially supported by the NSF CAREER award under Grant No. 1652515, the NSF grants OAC-1835712, OIA-1937043, CHS-1908767, CHS-1901091, NSERC DGECR-2021-00461 and RGPIN-2021-03707, a Sloan Fellowship, a gift from Adobe Research and a gift from Advanced Micro Devices, Inc.

\section*{Ethical approval}
All data used in this paper was provided, already anonymized, by 3Shape A/S. The data was originally acquired for diagnostic purposes unrelated to this study. No other aspect of this work triggered ethical issues.

\section*{Acknowledgements}
We thank NYU IT High-Performance. We also thank 3Shape A/S for providing this study's CBCT scans and, especially, the Dental CAD AI team for their support in the CBCT segmentation and computation of teeth axes. 

\section*{Declaration of Competing Interest}
Authors TG and PS are affiliated with 3Shape A/S, which provided support in the form of salaries and data for this study. Their commercial affiliation does not alter our adherence to sharing data and materials. Additionally, there are no patents, products in development, or marketed products associated with this research to declare.

\bibliographystyle{elsarticle-num} 
\bibliography{my_ref}

\end{document}